\documentclass{elsart}

\usepackage{graphicx}
\usepackage{longtable}
\usepackage{amssymb}
\usepackage{stmaryrd}
\begin{document}
\begin{frontmatter}
\title{Atomic mass measurements of short-lived nuclides around the doubly-magic $^{208}\mbox{Pb}$}
\author[1]{C.~Weber\corauthref{cor1}\thanksref{label1}\thanksref{label2}},
\ead{christine.weber@phys.jyu.fi}\corauth[cor1]{Corresponding
author.}
\thanks[label1]{This publication comprises part of the Ph.D. thesis of C. Weber.}
\thanks[label2]{Present address:
University of Jyv\"askyl\"a, FIN-40014 Jyv\"askyl\"a, Finland}
\author[2]{G.~Audi}, \author[1]{D.~Beck},
\author[1,3]{K.~Blaum\thanksref{label3}}, \thanks[label3]{Present
address: Universit\"at Mainz, D-55099 Mainz, Germany}
\author[4]{G.~Bollen}, \author[1,3]{F.~Herfurth},
\author[1]{A.~Kellerbauer\thanksref{label4}}, \thanks[label4]{Present
address: MPI f\"ur Kernphysik, D-69029 Heidelberg, Germany}
\author[1,5]{H.-J.~Kluge},
\author[2]{D.~Lunney}, and \author[4]{S.~Schwarz}
\address[1]{Gesellschaft f\"ur Schwerionenforschung mbH, D-64291 Darmstadt, Germany}
\address[2]{CSNSM-IN2P3/CNRS, Universit\'{e} de Paris-Sud, F-91405 Orsay, France}
\address[3]{CERN, CH-1211 Geneva 23, Switzerland}
\address[4]{NSCL, Michigan State University, East Lansing, MI 48824-1321, USA}
\address[5]{Universit\"at Heidelberg, D-69120 Heidelberg, Germany}
\begin{abstract}
Accurate atomic mass measurements of neutron-deficient and
neutron-rich nuclides around the doubly-magic $^{208}\mbox{Pb}$
and of neutron-rich cesium isotopes were performed with the
Penning trap mass spectrometer ISOLTRAP at ISOLDE/CERN. The masses
of $^{145,147}\mbox{Cs}$, $^{181,183}\mbox{Tl}$,
$^{186}\mbox{Tl}^{m}$, $^{187}\mbox{Tl}$, $^{196}\mbox{Tl}^{m}$,
$^{205}\mbox{Tl}$, $^{197}\mbox{Pb}^{m}$, $^{208}\mbox{Pb}$,
$^{190-197}\mbox{Bi}$, $^{209,215,216}\mbox{Bi}$,
$^{203,205,229}\mbox{Fr}$, and $^{214,229,230}\mbox{Ra}$ were
determined. The obtained relative mass uncertainty in the range of
$2 \cdot 10^{-7}$ to $2 \cdot 10^{-8}$ is not only required for
safe identification of isomeric states but also allows mapping the
detailed structure of the mass surface. A mass adjustment
procedure was carried out and the results included into the Atomic
Mass Evaluation. The resulting separation energies are discussed
and the mass spectrometric and laser spectroscopic data are
examined for possible correlations.
\end{abstract}

\end{frontmatter}
\section{\label{in}Introduction}
The accurate knowledge of atomic masses is required for many areas
of physics \cite{Lunn2003,Blau2006}. In particular, the mass gives
access to the nuclear binding energy which in turn reveals
distinct features of nuclear structure, notably shell effects and
their relative strength as a function of isospin. This paper
reports on mass data in the vicinity of the doubly-magic
$^{208}\mbox{Pb}$ nucleus. Strong nuclear structure effects are
observed in particular at the {\it neutron-deficient} side from $Z
= 77$ to $Z = 84$. They have been revealed by laser spectroscopic
measurements of the optical isotope shift
\cite{Otte1989,Klug2003}. Such measurements allow one to determine
the changes of the root-mean-square charge radii $\delta\langle
r^2 \rangle$ relative to a reference nucleus. Characteristic
irregularities were observed in the isotopic chains of iridium ($Z
= 77$), platinum ($Z = 78$), gold ($Z = 79$), mercury ($Z = 80$),
and thallium ($Z = 81$) at approximately mid-shell ($N = 104$).
The most prominent example is the odd-even shape staggering effect
within the mercury isotopes \cite{Bonn1972,Dabk1979}. These
phenomena were first explained for mercury as two different minima
in the potential energy surface at a weakly oblate and a stronger
prolate
deformation (shape coexistence) \cite{Frau1975}.\\
In addition, several rotational bands were observed in the nuclear
excitation spectra of these nuclei. An overview on nuclear shape
coexistence of even-mass nuclides is given in \cite{Wood1992}. In
$^{186}\mbox{Pb}$, even triple shape coexistence has been observed
\cite{Andr2000}. Here, the three lowest energy states having
spin/parity $0^+$ correspond to spherical, oblate and prolate
shapes, all within an energy spanning only $650~\mbox{keV}$. These
phenomena are described by particle-hole excitations of proton
pairs across the closed shell \cite{Heyd1983,Heyd1990}. The
interaction of additional proton-hole pairs and valence neutrons
will lead to a minimum in energy around mid-shell. If the energies
of the ground state and the intruder state are close, tiny changes
in the nuclear binding energy are decisive for the nuclear ground
state shape.\\
Such minute nuclear binding energy effects in this mass range ($A
\ge 180$) are only accessible by mass determinations with mass
uncertainties $\delta m$ of $10 - 20~\mbox{keV}$. This has been
demonstrated by ISOLTRAP measurements on neutron-deficient mercury
isotopes \cite{Schw1998,Schw2001,Foss2002}. In that work, isomeric
states, a 54-keV level in $^{187}\mbox{Hg}$ and a 128-keV level in
$^{191}\mbox{Hg}$ were even identified. This was the first example
of nuclear spectroscopy by mass spectrometry. The present work
complements the information on high-precision masses in this area
of the chart of nuclei with data on bismuth, lead, and thallium
isotopes. The interconnection of the nuclides studied in this work
by $\alpha$-decay chains results in an improvement of further mass
values up to $Z \approx 92$.\\
In case no direct experimental data existed, previous mass values
were obtained by a combination of up to four $Q_{\alpha}$ links
with results from mass spectrometry. For example, the
$\alpha$-decay of neutron-deficient bismuth was studied at the
LISOL facility for odd \cite{Coen1985} and even isotopes
\cite{VanD1991}. All direct data being available in this region
are either ISOLTRAP measurements \cite{Schw2001,Boll1992,Herf2005}
or mass determinations by the Schottky technique at the
Experimental Storage Ring ESR \cite{Rado2000,Litv2003,Litv2005}. A
detailed comparison to the latter data sets is made in Sec.
\ref{ESRcomp}. Mass values of neutron-rich cesium isotopes were
previously determined by triplet measurements with a
Mattauch-Herzog
spectrometer \cite{Ephe1979} and \cite{Audi1982}.\\
A particular challenge while addressing the neutron-deficient
nuclides experimentally is caused by the existence of two or even
three isomeric states with both short half-lives and low
excitation energies of sometimes less than $100~\mbox{keV}$. This
requires an extremely
high mass resolution of up to $10^7$.\\
Masses on the {\it neutron-rich} side of the doubly magic
$^{208}\mbox{Pb}$ nucleus are of great importance for predictions
on the stability of superheavy elements, which are stabilized by
shell corrections \cite{Paty1989}. A variation of this correction
by about $1~\mbox{MeV}$ leads to changes in the calculated
half-lives for spontaneous fission or $\alpha$ decay by several
orders of magnitude. Since no experimental mass values are
available in this region, model calculations of the shell
correction can be verified at the adjacent shell closure of  $Z =
82$.\\
Furthermore, accurate mass values are required in order to provide
reliable reference masses for the calibration of mass data
obtained in measurements at the Experimental Storage Ring ESR
\cite{Rado2000,Litv2003,Litv2005}. In an ESR mass determination on
neutron-deficient nuclides \cite{Litv2003,Litv2005} 117 reference
masses were used in order to calibrate the 466 mass values studied
in that experiment. In the neutron-rich region the number of
reference masses is particularly low.
\section{\label{ex}Experimental setup and procedure}
\begin{figure}
\centerline{\mbox{\includegraphics[width=0.7\textwidth]{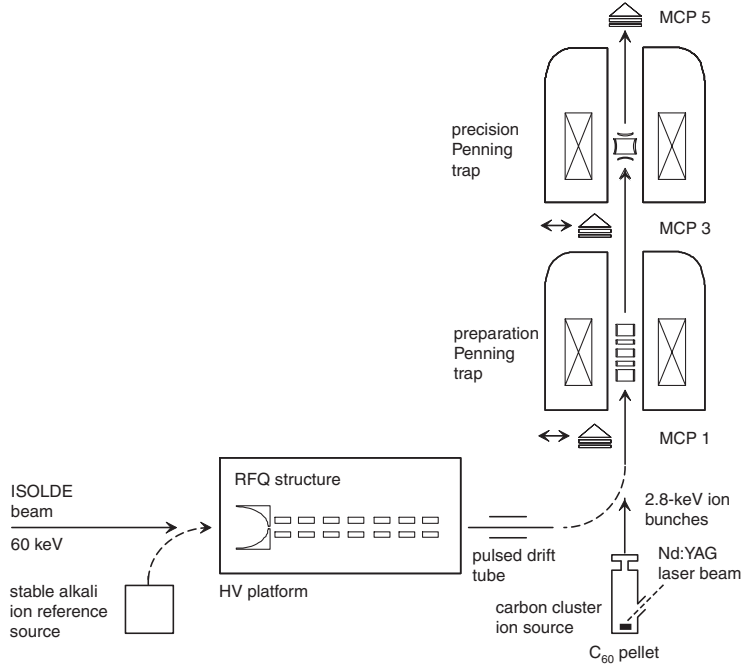}}}
\centerline{\parbox{0.75\textwidth}{\caption{\label{A_kap52_is_setup}Schematic
view of the Penning trap mass spectrometer ISOLTRAP (For details
see text).}}}
\end{figure}

\begin{figure}
\centerline{\mbox{\includegraphics[width=0.7\textwidth]{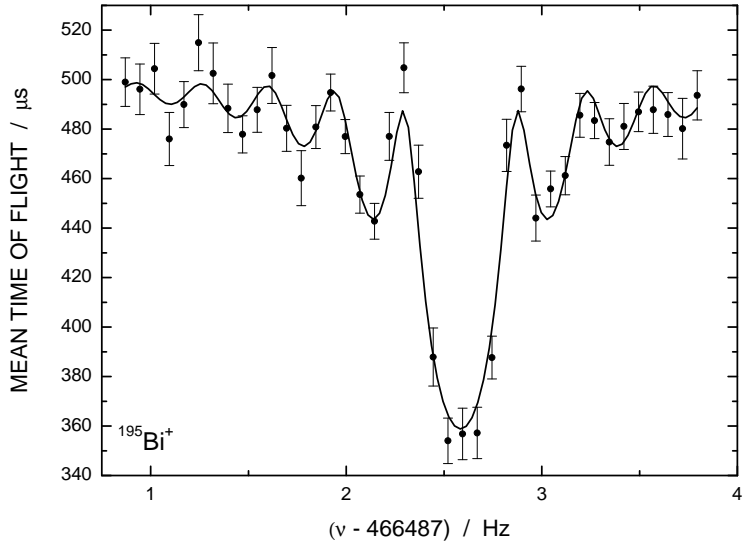}}}
\centerline{\parbox{0.75\textwidth}{\caption{\label{A_kap54_Zyklres_133}Cyclotron
resonance curve of $^{195}\mbox{Bi}^+$ ions. The solid line is a
fit of the theoretical lineshape \cite{Koni1995a} to the data
points.}}}
\end{figure}

The Penning trap mass spectrometer ISOLTRAP
\cite{Boll1996,Herf2001b,Mukh2007} is installed at the on-line
isotope separator ISOLDE \cite{Kugl2000} at CERN. Here,
mass-separated beams of short-lived nuclides are provided for
different experiments at an energy of $60~\mbox{keV}$. They are
produced by bombarding a thick target with high-energy
($1.4~\mbox{GeV}$) proton pulses ($3 \times 10^{13}$
protons/pulse) which induce spallation, fission, or fragmentation
reactions. Reaction products diffuse from the heated target into
an ion source region. Here, they are ionized either by surface
ionization, by electron impact in a hot plasma, or a resonant
photo ionization in a resonance ionization
laser ion source (RILIS) \cite{Mish1993,Kost2003}.\\
In this experiment nuclides around the shell closure of $Z = 82$
were produced by proton-induced (1.4-GeV) spallation and
fragmentation reactions on a $^{232}\mbox{ThC}_2$ target
($51~\mbox{g}/\mbox{cm}^{2}$). For lead and bismuth a resonant
laser ionization was applied, whereas cesium, thallium, francium,
and radium were surface-ionized. Finally, the ions were
mass-selected with a resolving power of $R = m/\Delta m
\mbox{(FWHM)} \approx 1000$ in the high-resolution separator
(HRS).\\
Figure \ref{A_kap52_is_setup} shows a schematic of the ISOLTRAP
mass spectrometer. The radiofrequency quadrupole (RFQ) cooler and
buncher \cite{Herf2001b} decelerates and accumulates the
continuous 60-keV ISOLDE beam and delivers a cooled, low-energy
ion bunch for efficient injection into the preparation Penning
trap \cite{Raim1997}.
Here, a mass-selective buffer-gas cooling scheme \cite{Sava1991}
is employed in order to separate isobars. A resolving power, $R =
\nu_{\mbox{\scriptsize c}} /\Delta \nu_{\mbox{\scriptsize c}}
\mbox{(FWHM)} = m/\Delta m \mbox{(FWHM)}$, of up to $10^5$ can be
achieved in this process. In the precision Penning trap, the mass
$m$ of a stored ion species with charge $q$ is determined via a
measurement of its cyclotron frequency $\nu_{\mbox{\scriptsize c}}
= qB/(2 \pi m)$, where $B$ denotes the strength of the magnetic
field. A cyclotron resonance spectrum is obtained by the
time-of-flight detection method \cite{Graf1980,Koni1995a}: The
extra radial kinetic energy resulting from the resonant excitation
with an azimuthal quadrupolar RF-field at $\nu_{\mbox{\scriptsize
c}}$ is detected by a reduction in the time of flight of the
ejected ions towards a detector (MCP5 in Fig.
\ref{A_kap52_is_setup}). The resolving power is approximately
given by the product of the cyclotron frequency and the
observation time: $R \approx 1.25 \cdot \nu_{\mbox{\scriptsize c}}
\cdot T_{\mbox{\scriptsize obs}}$ \cite{Boll2001}. A more detailed
description of a typical measurement cycle at ISOLTRAP can be
found in a recent publication
\cite{Sikl2005}.\\
An illustrative example of a cyclotron resonance curve is shown in
Fig. \ref{A_kap54_Zyklres_133}. In such a resonance curve, the
mean time of flight of the ions is displayed as a function of the
applied radiofrequency, and a resonant excitation can be
recognized by a shorter flight time. The diagram comprises 16
frequency scans containing a total amount of 1820 detected ions.
The linewidth of the central resonance, as given by the Fourier
limit $\Delta \nu_{\mbox{\scriptsize c}} \mbox{(FWHM)} \approx
1/T_{\mbox{\scriptsize obs}}$, results in a resolving power of $R
\approx 1.4 \times 10^{6}$ (mass resolution $\Delta m =
120~\mbox{keV}$) with an excitation time of the quadrupolar
RF-field $T_{\mbox{\scriptsize obs}} = 3~\mbox{s}$. This is
sufficiently high to resolve ground and isomeric states of
$^{195}\mbox{Bi}$. The theoretical lineshape \cite{Koni1995a} was
used to fit the data points. In order to calibrate the strength of
the magnetic field $B$, the cyclotron frequency of a reference
nuclide with well-known mass value is determined in regular time
intervals. This allows for the accurate mass determination of
short-lived nuclides far from stability with relative mass
uncertainties $\delta m/m \approx 10^{-8}$
\cite{Kell2003,Blau2003}. Examples of recent measurements on
$^{38}\mbox{Ca}$ \cite{Geor2007} as well as on several isotopes of
potassium \cite{Yazi2007}, nickel, copper, gallium
\cite{Guen2007}, and krypton \cite{Dela2006} cover a wide mass
range from $A = 35~\mbox{to}~95$. 
\section{\label{res}Data analysis}
\subsection{Determination of the cyclotron frequency ratios}
The primary result of an ISOLTRAP measurement is the cyclotron
frequency ratio $r = \nu_{\mbox{\scriptsize
c,ref}}/\nu_{\mbox{\scriptsize c}}$ of a reference ion with
well-known mass relative to the ion species to be determined.
Thus, these experimental ratios are independent from the
uncertainty of the reference mass $m_{\mbox{\scriptsize ref}}$.
From the ratios $r$, the atomic mass of the nuclide under study
can be deduced by
\begin{equation}
    \label{54-2}
    m = \frac{\nu_{\mbox{\scriptsize c,ref}}}
    {\nu_{\mbox{\scriptsize c}}}
    \left(m_{\mbox{\scriptsize ref}} - m_{\mbox{{\scriptsize e}}} \right) +
    m_{\mbox{{\scriptsize e}}}.
\end{equation}
Here, the atomic mass of the reference ion is indicated by
$m_{\mbox{{\scriptsize ref}}}$ and $m_{\mbox{{\scriptsize e}}}$ is
the mass of the electron. A correction for the binding energy of
the missing electron in the order of eV can be neglected.\\
Extensive studies using carbon cluster ions
\cite{Kell2003,Blau2002} helped to improve the assessment of
systematic effects. For example, systematic errors like
mass-dependent frequency shifts or the effect of a variation in
the magnetic field strength were quantified. In addition, these
investigations result in a better estimate of the final
uncertainty of the mass determination. The residual systematic
error of the ISOLTRAP mass measurements, estimated previously to
be $1 \times 10^{-7}$, was determined in \cite{Kell2003} to be
\begin{equation}
    \label{54-4}
     {\left(\frac{\delta m}{m}\right)}_{\mbox{\scriptsize res}}
        = 8 \times 10^{-9}.
\end{equation}
The data analysis was performed as described in \cite{Kell2003}.
Some aspects that are essential in the analysis of the data
presented here are summarized in the following.
\subsection{\label{counti}The presence of ions of different mass values}
Simultaneous storage of ions of different mass values may result
in a shift of the line center of the cyclotron frequency
resonance. Such possible contaminations depend on the ion
production process. In the case of measurements at ISOLTRAP,
isobars or nuclides in isomeric states might be present in
addition to the nuclide under investigation. Systematic
experimental studies using 10 to 70 ions stored in the trap as
well as simulations with two ions interacting by Coulomb forces
are discussed in \cite{Koni1991} and \cite{Boll1992a}: If the two
ions have different masses, the resulting frequency shifts depend
on the relative distance of the line centers divided by the
linewidth that is chosen during the measurement process. If both
lines cannot be resolved, only {\it one} cyclotron resonance is
observed at the frequency of the center of gravity. If the
cyclotron resonances of both ion species can be resolved, both
line centers are shifted to lower frequencies. The size of the
shift has been found to be
proportional to the number of stored ions.\\
To be able to exclude the possible presence of ions of different
mass in these measurements, the line center of the cyclotron
resonance is analyzed with respect to the number of ions
registered by MCP 5 (see Fig. \ref{A_kap52_is_setup}) in a
so-called count-rate-class analysis. The number of ions counted by
the microchannel plate detector after each irradiation of the
stored ion cloud with radiofrequency and ejection from the trap is
subdivided into at least three count-rate classes with comparable
average ion counts $N_{\mbox{\scriptsize{av}}}$. A possible shift
induced by a {\it contamination} is hence corrected by
extrapolating the cyclotron frequency to a vanishing number of
detected ions.
\begin{figure}%
\centerline{\mbox{\includegraphics[width=0.7\textwidth]{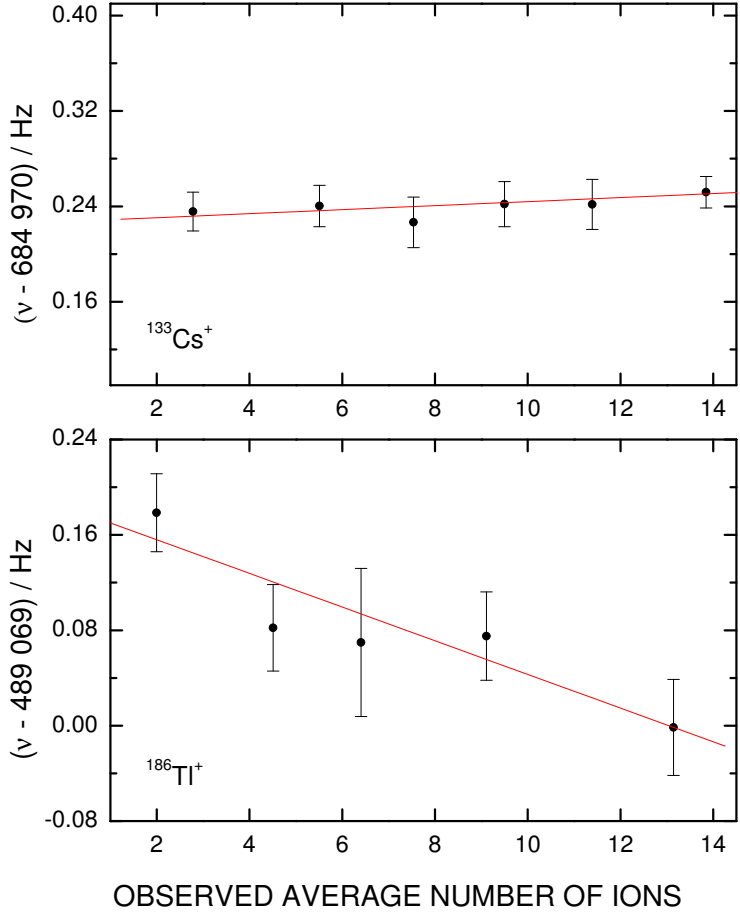}}}
\centerline{\parbox{0.75\textwidth}{\caption{\label{A_kap54_z-class}Dependance
of the frequency of the cyclotron resonance on the number of
detected ions for $^{133}\mbox{Cs}$ (top) and $^{186}\mbox{Tl}$
(bottom). Both measurements were performed with an excitation time
$T_{\mbox{\scriptsize{obs}}} = 1.5~s$. The center frequency $\nu$
is plotted versus the count-rate class containing an average
number of $N_{\mbox{\scriptsize{av}}} = 320~\mbox{ions}$ as
detected by MCP 5 (see Fig. \ref{A_kap52_is_setup}). The straight
line is a linear regression weighted by the individual error bars.
The frequency $\nu_{c}$, extrapolated to zero ions and its
respective error is used as the final result.}}}
\end{figure}
Two examples of a count-rate-class analysis, both with the same
average number of ions per count-rate class
$N_{\mbox{\scriptsize{av}}}$, are shown in Fig.
\ref{A_kap54_z-class}. The upper plot shows a measurement done
with stable $^{133}\mbox{Cs}$ ions from the off-line alkali ion
source (see Fig. \ref{A_kap52_is_setup}), which delivers only
surface-ionized alkali elements. The resulting fit of the slope of
$2(2)~\mbox{mHz/ion}$ shows no count-rate dependent effect. The
lower one shows a measurement of the radionuclide
$^{186}\mbox{Tl}$ as delivered by ISOLDE. With an increasing
number of detected ions per cycle a significant slope of
$-14(4)~\mbox{mHz/ion}$ of the cyclotron frequency towards lower
values is observed. This might be due to the presence of ions in
the isomeric state. See also the discussion on the nuclide
$^{186}\mbox{Tl}$ in Sec. \ref{evenA}. Combining several
measurements, the final value of a frequency ratio is obtained
from the weighted mean of individual results.
\subsection{\label{mass-dep}Correcting the mass-dependent systematic frequency shift}
An ion species $q/m$ experiences a frequency shift which can
originate for example from electric field imperfections or a
possible misalignment of the magnetic and electric field axis
versus each other \cite{Boll1996}. Since these shifts are to first
order {\it mass-independent}, this results in a systematic shift
in the determination of the frequency ratios, depending on the
mass difference $(m - m_{\mbox{\scriptsize{ref}}})$ between the
mass of interest {\it m} and the reference mass
$m_{\mbox{\scriptsize{ref}}}$. The upper limit of this effect has
been previously estimated to $\delta m/m \le 2 \times 10^{-9}
/{\mbox{u}}$ \cite{Beck1997b}. Frequency ratios between carbon
clusters whose values are exactly known were quantified by a
linear dependance \cite{Kell2003},
\begin{equation}
    \label{54-8}
     \frac{r - r_{\mbox{\scriptsize{th}}}}{r}
     = - 1.6(4) \cdot 10^{-10} \cdot (m - m_{\mbox{\scriptsize{ref}}}) /{\mbox{u}},
\end{equation}
where $r - r_{\mbox{\scriptsize{th}}}$ is the difference between
the measured and the true frequency ratio. In the experiments
reported here, covering the mass region between $A = 180$ and $A =
230$, the reference mass of
$^{133}\mbox{Cs}$\renewcommand\thefootnote{\fnsymbol{footnote}}\footnote{This
reference was chosen due to its high purity when delivered from
the ISOLTRAP off-line ion source.} ($\delta m = 25~\mbox{eV}$)
differs by about 70 mass units from the mass to be determined.
After applying the correction given by Eq. (\ref{54-8}) an
uncertainty of the same size,
\begin{equation}
    \label{54-9}
     \frac{\sigma_m(\nu_{\mbox{\scriptsize{ref}}})}{\nu_{\mbox{\scriptsize{ref}}}}
     =  1.6 \cdot 10^{-10} \cdot (m -
     m_{\mbox{\scriptsize{ref}}}) /{\mbox{u}},
\end{equation}
is added quadratically to the uncertainty. Due to the large mass
difference $(m - m_{\mbox{\scriptsize{ref}}})$, this yields an
additional error on the order of 1 keV to the deduced mass excess
value. In the end, the residual systematic uncertainty of $8\times
10^{-9}$ (Eq. (\ref{54-4})) is added quadratically to obtain the
final uncertainty.
\subsection{\label{assi}Mass-to-state assignment}
An unambiguous mass assignment to a specific state, ground state
or isomer, is not possible if a non-resolved mixture of nuclei in
their ground and isomeric states is present. If the production
ratio $R_{\mbox{\scriptsize{ig}}}$ (isomer to ground state), and
the isomeric excitation energy $E$ are known, the ground state
mass {\it m} can be extrapolated from the mass $m^*$,
corresponding to the observed cyclotron resonance, by
\cite{Audi1982}
\begin{equation}
    \label{54-10}
     m^* = m + \frac{R_{\mbox{\scriptsize{ig}}}}{R_{\mbox{\scriptsize{ig}}}+1} E.
\end{equation}
If the production ratio of two ion species in ground and isomeric
state is not known, the presence of the so-called contamination
can be observed in the data analysis by studying the behavior of
the resulting cyclotron frequency as a function of the count rate
class (see Fig. \ref{A_kap54_z-class}). Only in the rare case of
very similar half-lives of the ion under investigation and
contaminating ion,
contaminations might remain undiscovered.\\
Table \ref{A_kap5_NUBASE} lists all relevant data of the nuclides
that have been studied: Half-life $T_{1/2}$, spin state $I^{\Pi}$
and the isomeric excitation energy $E$. In case several isomeric
states are known to exist, those having a half-life longer than
$T_{1/2}~\ge~500~\mbox{ns}$ are
listed\renewcommand\thefootnote{\fnsymbol{footnote}}\footnote{Such
short half-lives are also given in order to provide information on
spin systematics. Nuclides with $T_{1/2} \le 50~\mbox{ms}$ are not
considered as contaminants in the studies reported here, since the
typical storage time of $1~\mbox{s}$ exceeds by far $T_{1/2}$.}.
The assignment of a measured mass value to an isomeric state is
performed under the assumption that isomeric states with a given
$I^{\Pi}$ and excitation energy might appear as listed in Tab.
\ref{A_kap5_NUBASE}. The chosen RF-excitation time results in a
width of the cyclotron resonance $\Delta \nu_{\mbox{\scriptsize
c}} \mbox{(FWHM)}$, which corresponds to a width $\Delta m
\mbox{(FWHM)} c^2$ in energy units. This value is also called mass
resolution and is listed in the last column, since it serves as an
important criterion for the observation of a contamination and
hence for the identification of the states.

\begin{center}
\begin{longtable}{lrrlccc}
\caption{\label{A_kap5_NUBASE} Half-lives {$T_{1/2}$}, spin states
$I^{\Pi}$ and excitation energies $E$ of the nuclides studied in
the present work (from NUBASE \cite{Audi1997} and
\cite{Audi2003a}). The full width at half maximum of the cyclotron
resonance in energy units $\Delta mc^2$ that has been used in
individual measurements is listed in the last column. Explanations
are given at the end of the table.}
\\\hline
Nuclide & &\multicolumn{2}{c}{Half-Life} & Spin/Parity & Excitation & $\Delta m \mbox{(FWHM)} c^2$\\
        & &\multicolumn{2}{c}{$T_{1/2}$} & $I^{\Pi}$    & energy $E$ [keV] &   [keV]\\
\hline
\endfirsthead
\caption{(Continued).}
\\\hline
Nuclide & &\multicolumn{2}{c}{Half-Life} & Spin/Parity  & Excitation & $\Delta m \mbox{(FWHM)} c^2$\\
        & &\multicolumn{2}{c}{$T_{1/2}$} & $I^{\Pi}$     & energy $E$ [keV] &  [keV]\\
\hline
\endhead
\hline
\endfoot
\multicolumn{7}{l} {
\begin{minipage}[t]{\textwidth}
\small
\begin{tabbing}
\=Isomeric assignment:
\\
\>$*$ \hspace*{0.5cm} \=\begin{minipage}[t]{8.35cm}In case the
uncertainty $\sigma$ of the excitation energy $E$ is larger than
half the energy ($\sigma \ge E / 2$) an asterisk has been added.
\end{minipage}\\[4pt]
\>\&      \>\begin{minipage}[t]{8.35cm}In case the ordering of the
ground and isomeric states have been reversed compared to ENSDF
(Evaluated Nuclear Structure Data Files), an '\&' sign has been
added.
\end{minipage}\\[5pt]
\>Half-life:\\
\>\# \>\begin{minipage}[t]{8.35cm}values estimated from systematic
trends in neighboring nuclides with the same {\it Z} and {\it N}
parities (even or odd).
\end{minipage}\\[4pt]
\>Ey \>\begin{minipage}[t]{8.35cm}$1~\mbox{Ey}= 10^{18}~\mbox{y}$,
1 exayear.
\end{minipage}\\[5pt]
\>Spin and parity:\\
\>( )  \>uncertain spin and/ or parity.\\[4pt]
\>\# \>\begin{minipage}[t]{8.35cm}values estimated from systematic
trends in neighboring nuclides with the same {\it Z} and {\it N}
parities (even or odd).
\end{minipage}\\[4pt]
\end{tabbing}
\normalsize
\end{minipage}
}
\endlastfoot
    $^{145}\mbox{Cs}$      &     & 582       & ms            & $3/2^{+}$         &               & 220\\
    $^{147}\mbox{Cs}$      &     & 225       & ms            & ($3/2^{+}$)       &               & 660\\
    $^{181}\mbox{Tl}$      &     & 3.4       & s             & $1/2^{+}$\#       &               & 340\\
    $^{181}\mbox{Tl}^{m}$  &     & 2.7       & ms            & $9/2^{-}$\#       & 850(30)       & 340\\
    $^{183}\mbox{Tl}$      &     & 6.9       & s             & $1/2^{+}$\#       &               & 340\\
    $^{183}\mbox{Tl}^{m}$  &     & 60        & ms            & $9/2^{-}$\#       & 625(17)       & 340\\
    $^{186}\mbox{Tl}$      & *\& & 40\#      & s             & ($2^{-}$)         &               & 210\\
    $^{186}\mbox{Tl}^{m}$  & *\& & 27.5      & s             & ($7^{+}$)         & 40(190)       & 210\\
    $^{186}\mbox{Tl}^{n}$  &     & 2.9       & s             & ($10^{-}$)        & 420(190)      & 210\\
    $^{187}\mbox{Tl}$      &     & $~$51     & s             & ($1/2^{+}$)       &               & 540\\
    $^{187}\mbox{Tl}^{m}$  &     & 15.6      & s             & ($9/2^{-}$)       & 332(4)        & 540\\
    $^{196}\mbox{Tl}$      &     & 1.84      & h             & $2^{-}$           &               & 120\\
    $^{196}\mbox{Tl}^{m}$  &     & 1.41      & h             & ($7^{+}$)         & 394.2(5)      & 120\\
    $^{205}\mbox{Tl}$      &     &\multicolumn{2}{c}{stable} & $1/2^{+}$         &               & 430\\
    $^{197}\mbox{Pb}$      &     & 8         & min           & $3/2^{-}$         &               & 120\\
    $^{197}\mbox{Pb}^{m}$  &     & 43        & min           & $13/2^{+}$        & 319.3(7)      & 120\\
    $^{208}\mbox{Pb}$      &     &\multicolumn{2}{c}{stable} & $0^{+}$           &               & 270\\
    $^{190}\mbox{Bi}$      & *   &  6.3      & s             & ($3^{+}$)         &               & 560\\
    $^{190}\mbox{Bi}^{m}$  & *   &  6.2      & s             & ($10^{-}$)        & 150(190)      & 560\\
    $^{190}\mbox{Bi}^{n}$  &     & $\ge$ 500 & ns            & $7^{+}$\#         & 420(190)      & 560\\
    $^{191}\mbox{Bi}$      &     & 12.3      & s             & ($9/2^{-}$)       &               & 110\\
    $^{191}\mbox{Bi}^{m}$  &     & 128       & ms            & ($1/2^{+}$)       & 242(7)        & 110\\
    $^{192}\mbox{Bi}$      &     & 34.6      & s             & ($3^{+}$)         &               & 110\\
    $^{192}\mbox{Bi}^{m}$  &     & 39.6      & s             & ($10^{-}$)        & 210(50)\#     & 110\\
    $^{193}\mbox{Bi}$      &     & 67        & s             & ($9/2^{-}$)       &               & 110\\
    $^{193}\mbox{Bi}^{m}$  &     & 3.2       & s             & ($1/2^{+}$)       & 308(7)        & 110\\
    $^{194}\mbox{Bi}$      & *   & 95        & s             & ($3^{+}$)         &               & 580\\
    $^{194}\mbox{Bi}^{m}$  & *   & 125       & s             & ($6^{+}$,$7^{+}$) & 100(70)\#     & 580\\
    $^{194}\mbox{Bi}^{n}$  & *   & 115       & s             & ($10^{-}$)        & 230(310)      & 580\\
    $^{195}\mbox{Bi}$      &     & 183       & s             & $9/2^{-}$\#       &               & 120\\
    $^{195}\mbox{Bi}^{m}$  &     & 87        & s             & $1/2^{+}$\#       & 399(6)        & 120\\
    $^{196}\mbox{Bi}$      &     & 5.1       & min           & ($3^{+}$)         &               & 590\\
    $^{196}\mbox{Bi}^{m}$  &     & 0.6       & s             & ($7^{+}$)         & 167(3)        & 590\\
    $^{196}\mbox{Bi}^{n}$  &     & 4         & min           & ($10^{-}$)        & 270(4)        & 590\\
    $^{197}\mbox{Bi}$      &     & 9.3       & min           & ($9/2^{-}$)       &               & 600\\
    $^{197}\mbox{Bi}^{m}$  &     & 5.04      & min           & ($1/2^{+}$)       & 350(160)      & 600\\
    $^{209}\mbox{Bi}$      &     & 19        & Ey            & $9/2^{-}$         &               & 270\\
    $^{215}\mbox{Bi}$      &     & 7.6       & min           & ($9/2^{-}$)       &               & 470\\
    $^{215}\mbox{Bi}^{m}$  &     & 36.4      & min           & ($25/2^{-}$)      & 1347.5(2.5)   & 470\\
    $^{216}\mbox{Bi}$      &     & 2.17      & min           & $1^{-}$\#         &               & 480\\
    $^{203}\mbox{Fr}$      &     & 550       & ms            & $9/2^{-}$\#       &               & 420\\
    $^{205}\mbox{Fr}$      &     & 3.85      & s             & ($9/2^{-}$)       &               & 260\\
    $^{229}\mbox{Fr}$      &     & 50.2      & s             & $1/2^{+}$\#       &               & 810\\
    $^{214}\mbox{Ra}$      &     & 2.46      & s             & $0^{+}$           &               & 470\\
    $^{229}\mbox{Ra}$      &     & 4         & min           & $5/2^{(+)}$       &               & 810\\
    $^{230}\mbox{Ra}$      &     & 93        & min           & $0^{+}$           &               & 330\\
    \hline
\end{longtable}
\end{center}
%
\subsection{Effects determining the production ratio of isomeric states}
In the measurements reported here possible contaminations are
mostly isomeric states, which could not be separated in the
preparation Penning trap. Exceptions are the heavy isobars
$^{229}\mbox{Fr}$ and $^{229}\mbox{Ra}$. The spallation process
does not transfer large angular momentum and hence low-spin states
are predominantly produced. Fragmentation reactions lead to a
production of higher spin states. Nuclides in states with
half-lives shorter than few 10~ms are suppressed due to the slow
release time of the reaction products from the ISOLDE target and
the time needed for the isobaric separation in the preparation
Penning trap (see Fig. \ref{A_kap52_is_setup}).\\
A further selection of an atom in a specific nuclear state can be
obtained by resonance laser ionization (RIS) with the ISOLDE laser
ion source and a proper choice of laser wavelengths and bandwidths
\cite{Mish1993}. Due to different hyperfine splittings and isotope
shifts in the ground and isomeric state, RIS can enhance the selectivity.\\
In the case of bismuth investigated in the present work, the
lasers were used in the broad-band mode. The production ratios in
this experiment can not be deduced from previous measurements with
the conventional ion source at ISOLDE. For lead, the applied
broad-band mode (laser bandwidth = 10 GHz) does not result in a
selection of isomeric states, since the isomeric shift and
hyperfine splitting are smaller or of the same size as the laser
bandwidth.
\section{\label{cyc-ratios}Experimental results}
Table \ref{A_kap5_ratios} lists all frequency ratios that were
determined relative to the cyclotron frequency of
$^{133}\mbox{Cs}$ ions. If possible, the assignment to a
particular isomeric state is indicated and will be discussed in
Sec. \ref{assign}. The last column lists all mass excess values
deduced from the determined frequency ratios $r$.
\begin{center}
\begin{longtable}{llcr}
\caption{\label{A_kap5_ratios}Cyclotron frequency ratios {\it r}
of the $^{133}\mbox{Cs}^+$ reference ion with
$\nu_{\mbox{\scriptsize{c,ref}}}$ relative to the cyclotron
frequency $\nu_{\mbox{\scriptsize{c}}}$ of the nuclide under
study. The indices {\it m} and {\it x} behind the element symbol
indicate the first excited isomeric state (m) or a possibly
unresolved mixture of different states (x). The third column gives
the relative total uncertainty $\sigma_{\mbox{\scriptsize{rel}}}$.
The last column lists the mass excess as derived from
$ME~=~[m~{\mbox{(in~u)} - A}]~{\mbox{(in~keV)}}$ using the
reference mass $m_{\mbox\scriptsize{ref}}$ of $^{133}\mbox{Cs}$
ions \cite{Brad1999}.}
\\\hline
Nuclide & Frequency ratio  & $\sigma_{\mbox{\scriptsize{rel}}}$ &{Mass excess} \\
        & ~~~~$r = \frac {\nu_{\mbox{\scriptsize{c,ref}}}}{\nu_{\mbox{\scriptsize{c}}}}$  & & [keV]~~~~ \\
\hline
\endfirsthead
\caption{(Continued).}
\\\hline
Nuclide & Frequency ratio  & $\sigma_{\mbox{\scriptsize{rel}}}$ &{Mass excess} \\
        & ~~~~$r = \frac {\nu_{\mbox{\scriptsize{c,ref}}}}{\nu_{\mbox{\scriptsize{c}}}}$  & & [keV]~~~~ \\
\hline
\endhead
$^{145}\mbox{Cs}$       &   1.090516444(91)   &      $8.3 \times 10^{-8} $        &   -60052(11)   \\
$^{147}\mbox{Cs}$       &   1.10562975(48) &      $4.3 \times 10^{-7} $        &   -52011(60)   \\
$^{181}\mbox{Tl}$       &   1.361768475(82)   &      $6.0 \times 10^{-8} $        &   -12802(10)   \\
$^{183}\mbox{Tl}$       &   1.376786217(81)   &      $5.9 \times 10^{-8} $        &   -16592(10)   \\
$^{186}\mbox{Tl}^{m}$   &   1.399332234(69)   &      $5.0 \times 10^{-8} $        &   -19874.4(8.6)\\
$^{187}\mbox{Tl}^{x}$   &   1.40683800(18) &      $1.3 \times 10^{-7} $        &   -22154(23)   \\
$^{196}\mbox{Tl}^{m}$   &   1.474515617(98)   &      $6.7 \times 10^{-8} $        &   -27103(12)   \\
$^{205}\mbox{Tl}$       &   1.542259739(82)   &      $5.3 \times 10^{-8} $        &   -23818(10)   \\
$^{197}\mbox{Pb}^{m}$   &   1.482061386(45)   &      $3.0 \times 10^{-8} $        &   -24429.5(5.5)\\
$^{208}\mbox{Pb}$       &   1.564849038(42)   &      $2.7 \times 10^{-8} $        &   -21742.2(5.2)\\
$^{190}\mbox{Bi}^{x}$   &   1.42950438(20) &      $1.4 \times 10^{-7} $        &   -10535(25)   \\
$^{191}\mbox{Bi}$       &   1.437006669(64)   &      $4.5 \times 10^{-8} $        &   -13244.9(8.0)  \\
$^{192}\mbox{Bi}^{m}$   &   1.444529605(72)   &      $5.0 \times 10^{-8} $        &   -13398.5(9.0)  \\
$^{193}\mbox{Bi}$       &   1.452033781(84)   &      $5.8 \times 10^{-8} $        &   -15875(10)   \\
$^{194}\mbox{Bi}^{m}$   &   1.45955793(12) &      $8.0 \times
10^{-8} $        &   -15878(50)\footnote{The original error of
$14~\mbox{keV}$ is increased in the AME2003 due to a possible
contamination by
the $3^+$ and $10^-$ states.}   \\
$^{195}\mbox{Bi}$       &   1.467064775(45)   &      $3.1 \times 10^{-8} $        &   -18023.7(5.6)\\
$^{196}\mbox{Bi}^{m}$   &   1.47459021(12) &      $7.9 \times
10^{-8} $        &   -17868(50)~${^{**}}$   \\
$^{197}\mbox{Bi}^{x}$   &   1.482099666(89)   &      $6.0 \times
10^{-8} $        &   -19706(25)\footnote{Value corrected by $-16(22)~\mbox{keV}$ in the AME2003 for possible contamination from $^{197}\mbox{Bi}^{m}$.}\\
$^{209}\mbox{Bi}$       &   1.572401384(35)   &      $2.2 \times 10^{-8} $        &   -18254.9(4.3)\\
$^{215}\mbox{Bi}$       &   1.61770721(12) &      $7.3 \times 10^{-8} $        &   1648(15)     \\
$^{216}\mbox{Bi}$       &   1.625265520(93)   &      $5.7 \times 10^{-8} $        &   5873(11)     \\
$^{203}\mbox{Fr}$       &   1.52741073(13) &      $8.6 \times 10^{-8} $        &   861(16)      \\
$^{205}\mbox{Fr}$       &   1.542441561(73)   &      $4.7 \times 10^{-8} $        &   -1308.6(9.0)   \\
$^{229}\mbox{Fr}$       &   1.72332168(30) &      $1.7 \times 10^{-7} $        &   35816(37)    \\
$^{214}\mbox{Ra}$       &   1.61017053(17) &      $1.0 \times 10^{-7} $        &   100(20)      \\
$^{229}\mbox{Ra}$       &   1.72329550(16) &      $9.1 \times 10^{-8} $        &   32575(19)    \\
$^{230}\mbox{Ra}$       &   1.73083537(10)  &      $5.8 \times 10^{-8} $        &   34518(12)    \\
\hline
\end{longtable}
\end{center}
%
\section{\label{AME}Atomic mass evaluation}
\begin{figure}
\centerline{\mbox{\includegraphics[width=0.8\textwidth]{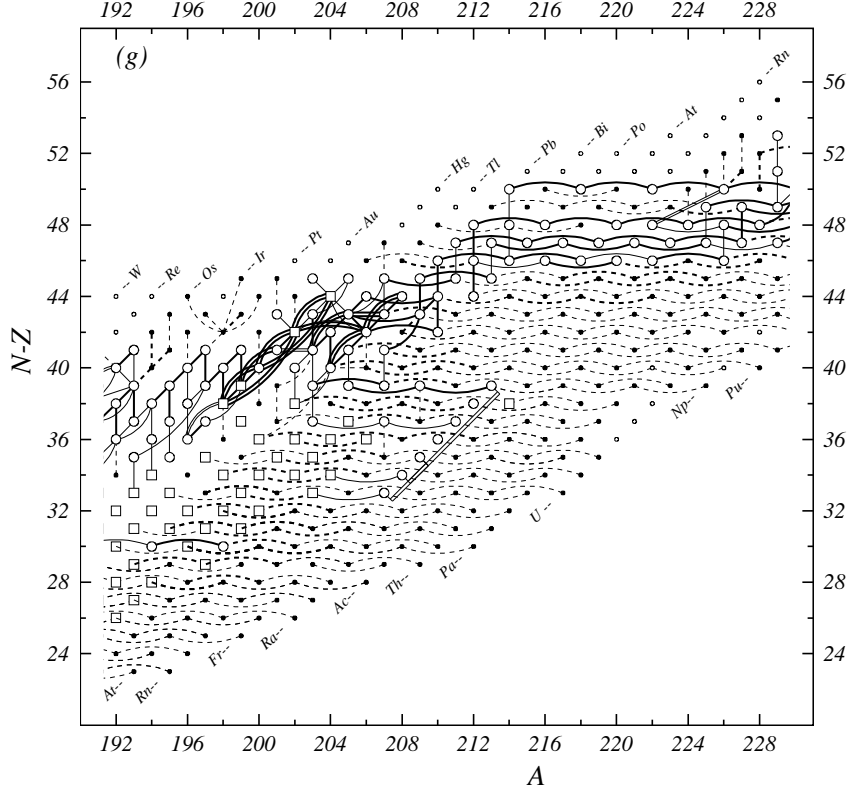}}}
\centerline{\parbox{0.9\textwidth}{\caption{\label{connect}Connection
diagram between nuclides of mass number {\it A} and neutron excess
({\it N} - {\it Z} ). The symbols indicate primary nuclides (big
symbols, with '$\boxempty$' being an absolute mass-doublet
nuclide) and secondary nuclides (small symbols). The latter ones
can be distinguished between experimental data (filled) and
extrapolated values from systematic trends (open). Primary
connections are direct mass spectroscopic data (double line) or
reaction data (single line), while thicker lines indicate data
with higher precision. All secondary nuclides are connected via
dashed lines. This plot is taken from the AME2003
\cite{Waps2003}.}}}
\end{figure}

The Atomic Mass Evaluation AME is an adjustment procedure giving
mass data for all those nuclides where relevant experimental
information, like for example mass values or half-lives, are
known. Such an evaluation was performed in December 2002 before
these ISOLTRAP data were included. This atomic mass evaluation is
referred to in the following as Intermediate Atomic Mass
Evaluation 2002 (IAME2002). It is an up-to-date adjustment after
the published issue of AME1995 \cite{Audi1995} and the NUBASE
compilation \cite{Audi1997}. The assignment of the observed
cyclotron resonances to nuclear states was not straightforward at
all since this area of the chart of nuclei is rich in low-lying
isomers. Subsequently, the ISOLTRAP results, reported in this
publication, were included into the Atomic Mass Evaluation 2003. A
detailed description of the adjustment procedure is given in the
particular issues of the AME \cite{Audi1995,Audi1993,Audi2003b}.
Here, only some general features that are relevant for this work
are described. As a result of the AME, adjusted mass values are
obtained from different types of experimental input data. Even
though a possible existence of more than 6000 different
combinations of protons and neutrons is predicted, data from only
approximately 2830 nuclides are available \cite{Audi2001}. If
direct experimental data are not existing for nuclides near to the
region of known masses, mass values are extrapolated from
systematic, experimental trends. All experimental results can be
represented as a relation between two (and more) nuclides. They
span a net of interconnections between nuclides across the whole
nuclear chart as it can be seen in the diagram for $A=192$ to
$A=229$ (Fig. \ref{connect}).
For the present ISOLTRAP data giving a frequency ratio to a
nuclide with precisely known atomic mass value, these connections
are linked to the mass of $^{133}\mbox{Cs}$, which has been
determined with a relative uncertainty of only $0.2~\mbox{ppb}$
\cite{Brad1999}.\\
All input data are transformed into a set of linear equations. For
the case of experimental frequency ratios this is explained in
\cite{Beck2000}. Each datum $x_{i}$ with a standard deviation
$dx_{i}$ represents a link between two, three, or four unknown
masses $m_{\lambda}$. The over-determined set of {\it X} equations
for {\it Y} mass values $(X > Y)$ is represented as a set of
linear equations with {\it Y} parameters and is finally solved by
the least-squares method as described in \cite{Audi2001}. Most of
the ISOLTRAP frequency ratios that are published in the present
work were introduced already into the most recently published
atomic mass evaluation, {\it i.e.} AME2003 \cite{Audi2003b}.
\section{\label{assign}Assignment of the observed cyclotron resonances to nuclear states}
The assignment of a measured mass value to an isomeric state is
performed under the assumption that isomeric states with a given
$I^{\Pi}$ and excitation energy might appear as listed in Tab.
\ref{A_kap5_NUBASE}.
\subsection{Nuclides with odd mass number $A$}
\begin{figure}
\centerline{\mbox{\includegraphics[width=0.7\textwidth]{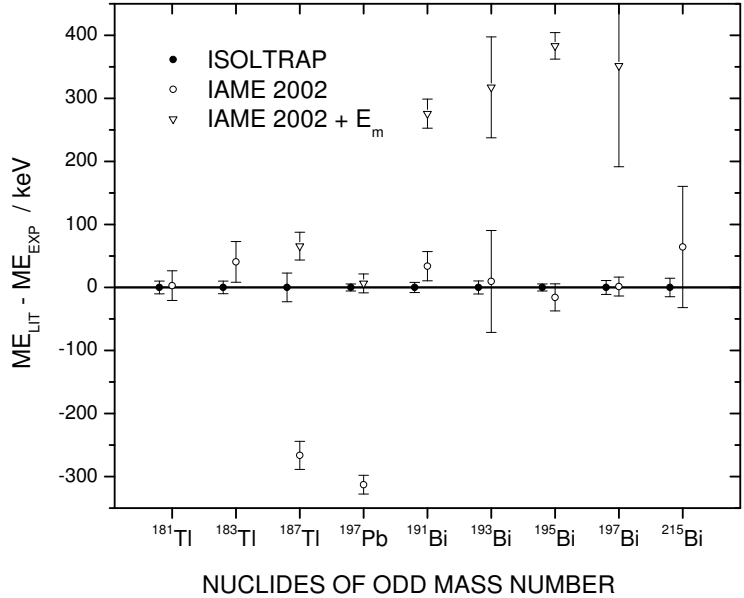}}}
\centerline{\parbox{0.75\textwidth}{\caption{\label{odd}Comparison
of ISOLTRAP mass values for odd-even and even-odd nuclides to
literature values. The zero line represents the experimental
ISOLTRAP data with their uncertainties. Open symbols indicate the
results of an intermediate atomic mass evaluation performed end of
2002 (IAME2002). IAME2002 includes all literature values available
at that time but not the ISOLTRAP data. The excitation energies
$E_{\mbox{\scriptsize{m}}}$ are taken from NUBASE
\cite{Audi1997,Audi2003a}.}}}
\end{figure}

In this section those odd-mass nuclides having a long-lived
isomeric state are discussed. According to Tab.
\ref{A_kap5_NUBASE} the ground states of $^{181, 183,
187}\mbox{Tl}$ and $^{197}\mbox{Pb}$ have always a low spin value,
whereas in the bismuth isotopes $^{191,193,195,197}\mbox{Bi}$ the
ground state has a high spin.\\
The comparison of the results of Tab. \ref{A_kap5_ratios} with
those of the intermediate adjustment (IAME2002) \cite{Audi2003a},
helps to indicate which spin states have most probably been
produced during the experiment. Shown in Fig. \ref{odd} is such a
comparison of the ISOLTRAP masses to the IAME2002 masses with and
without the added isomeric energy. The zero line depicts the
experimental ISOLTRAP data with their uncertainties. The
differences to the literature values are indicated by open
symbols, whereas the position of excited nuclear states is taken
from the current NUBASE data (Tab. \ref{A_kap5_NUBASE})
\cite{Audi1997,Audi2003a}. The uncertainty of the excited state is
derived by adding the uncertainties of the
ground state mass value and the excitation energy quadratically.\\

{\bf $^{181}\mbox{Tl}$} and {\bf $^{183}\mbox{Tl}$}: Here, the
half-lives of the excited isomeric states ($I^\pi = 9/2^+$) of
$2.7~\mbox{ms}$ for $^{181}\mbox{Tl}$ and $60~\mbox{ms}$ for
$^{183}\mbox{Tl}$, are much shorter than the duration of a
measurement cycle (approximately $\!1~\mbox{s}$). Therefore, the
measured cyclotron frequency ratios are clearly assigned to the
respective ground states. The measurements of ISOLTRAP agree with
the values from the intermediate adjustment IAME2002 (compare Fig.
\ref{odd}), but the uncertainties got significantly
improved.\\

{\bf $^{187}\mbox{Tl}$}: The measurement of the nuclide
$^{187}\mbox{Tl}$ has been performed with a mass resolution of
$540~\mbox{keV}$, which is insufficient to resolve both ground and
isomeric states, differing by $E = 332~\mbox{keV}$. Hence, an
eventually present admixture of either of the states can not be
excluded. The determined ISOLTRAP value suggests a predominant production of the isomer.\\

{\bf $^{197}\mbox{Pb}^{m}$}: The mass values of both ground and
isomeric states in $^{197}\mbox{Pb}$ are listed in IAME2002 with
uncertainties of only $15~\mbox{keV}$. The value resulting from
the ISOLTRAP measurement with a final uncertainty of
$5.5~\mbox{keV}$ agrees within the error bars to the one of the
excited $I^{\Pi} = 13/2^+$ state ($E = 319.3~\mbox{keV}$) and
hence is assigned to it.\\

{\bf $^{191,193,195,197}\mbox{Bi}$, $^{215}\mbox{Bi}$}: A series
of bismuth isotopes with odd mass numbers $A = 191~\mbox{to}~197$
and $A = 215$ has been studied. With the exception of the case of
$^{197}\mbox{Bi}$ the mass resolution $\Delta m$ (see Tab.
\ref{A_kap5_NUBASE}) is at least a factor of two smaller than the
mass difference between both states. Therefore, a contamination of
the respective second state is definitely excluded in the
count-rate-class analysis. The remaining question which of the two
states has been measured cannot be solved easily. Yield
measurements for lead and bismuth isotopes at ISOLDE reveal that
states having higher spin values are predominantly produced
\cite{Kost2003}. These are for $191 \le A \le 197$ the $I^{\Pi} =
9/2^-$ ground states of bismuth. Indeed, the comparison of the
ISOLTRAP results shows a good agreement with the values of the
intermediate evaluation IAME2002 for the ground states of
$^{191,193,195,197}\mbox{Bi}$. The measurements of
$^{197}\mbox{Bi}$ were performed with a resolution of
$600~\mbox{keV}$ which is not sufficient to resolve ground and
isomeric state, differing by $350(160)~\mbox{keV}$. Measurements
of the neighboring odd-$A$ isotopes $^{193}\mbox{Bi}$ and
$^{195}\mbox{Bi}$ with approximately $115~\mbox{keV}$ resolution
did not show any hint of a contamination by an excited state. A
systematic comparison of individual measurements with low
($580~\mbox{keV}$) and high ($115~\mbox{keV}$) resolution was used
in order to quantify a correction by a possible contamination by
the excited $^{197}\mbox{Bi}^{m}$ state. This correction of
$-16(22)~\mbox{keV}$ increases the uncertainty of the ISOLTRAP
value in the atomic mass table AME2003 from $11~\mbox{keV}$ to
$25~\mbox{keV}$.\\
Considering the isomeric states of all odd-mass nuclides studied
here a consistent picture is observed. Except for cases where the
presence of a state can be excluded right away due to its short
half-life, higher spin states are predominantly produced. In
$^{187}\mbox{Tl}$ and $^{197}\mbox{Pb}$ these are the {\it
excited} states with spins of $9/2^-$ and $13/2^+$. In bismuth ($A
\le 197$) the ground states have a higher spin value of $9/2^-$.
For the nuclide $^{215}\mbox{Bi}$ with an excited state of $>
1~\mbox{MeV}$ energy and a spin $25/2^-$, the $9/2^-$ ground state
mass value has been determined.

\subsection{\label{evenA}Nuclides with even mass number $A$}
\begin{figure}
\centerline{\mbox{\includegraphics[width=0.7\textwidth]{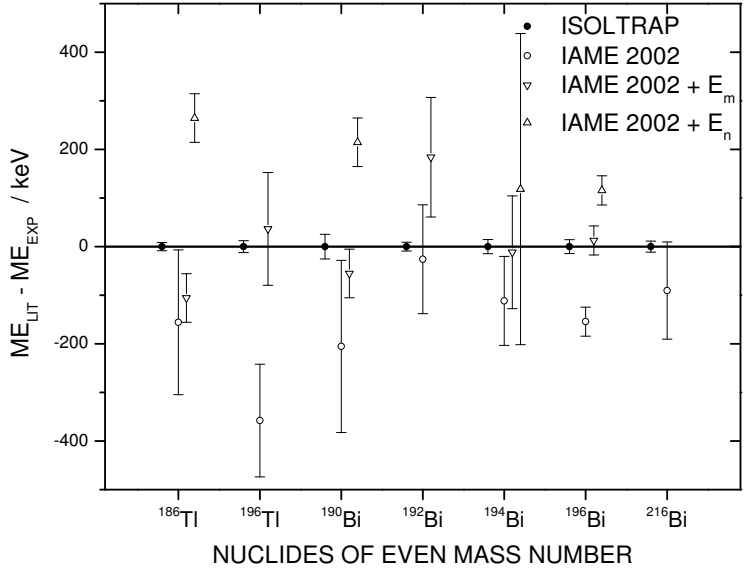}}}
\centerline{\parbox{0.75\textwidth}{\caption{\label{even}Comparison
of even-$A$ mass values from ISOLTRAP to an intermediate atomic
mass evaluation (IAME2002). The zero line represents the
experimental data points obtained within this work with their
uncertainties. Excited, long-lived states as they are listed in
NUBASE \cite{Audi1997,Audi2003a} are shown by open triangles.
Those isomeric states with the higher excitation energy
$E_{\mbox{\scriptsize{n}}}$ are depicted by a triangle pointing
upwards.}}}
\end{figure}

In order to facilitate a spin-to-state assignment, the present
ISOLTRAP data are plotted in Fig. \ref{even} and compared relative
to ground state mass values and excitation energies from the
intermediate evaluation IAME2002. These nuclides have sometimes
{\it two} excited, long-lived states (see also Tab.
\ref{A_kap5_NUBASE}). Excitation energies of states with
half-lives longer than $500~\mbox{ns}$ are shown.\\
Apart from the $500~\mbox{ns}$-state of $^{190}\mbox{Bi}$ all
nuclides have half-lives that are sufficiently long ($\ge
50~\mbox{ms}$) in order to be studied with the Penning trap mass
spectrometer ISOLTRAP. Due to the lack of any quantitative
information on production ratios, a conclusion can be only based
on a systematic comparison within the current data. In analogy to
the nuclides of odd mass number, it is expected that lower spin
states will be produced with lower yields.\\

\begin{figure}
\centerline{\mbox{\includegraphics[width=0.7\textwidth]{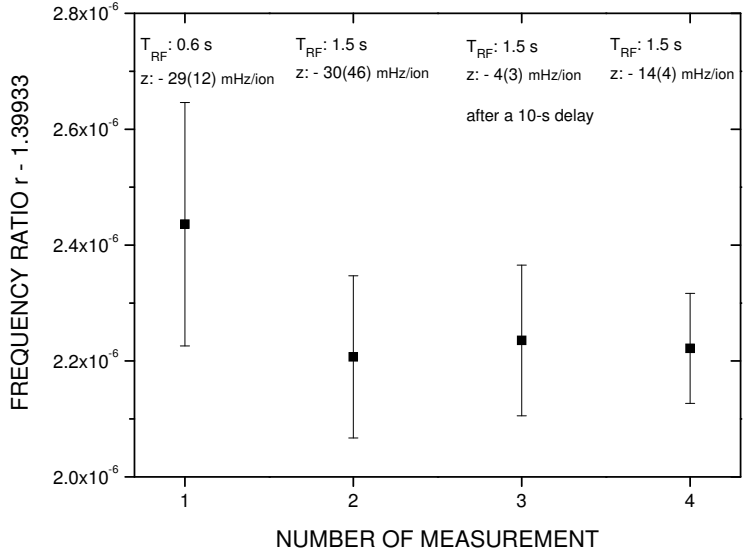}}}
\centerline{\parbox{0.75\textwidth}{\caption{\label{Tl186}Four
cyclotron frequency measurements for $^{186}\mbox{Tl}$ with two
different excitation times $T_{\mbox{\scriptsize{RF}}}$ as
indicated. $T_{\mbox{\scriptsize{RF}}} = 0.6~\mbox{s}$ corresponds
to a mass resolution of $\Delta m = 530~\mbox{keV}$ and
$T_{\mbox{\scriptsize{RF}}} = 1.5~\mbox{s}$ to $\Delta m =
210~\mbox{keV}$. The result of the count-rate-class analysis of
the cyclotron resonance of $^{186}\mbox{Tl}$ is given for each
data point. In the measurement \#3 an additional waiting time in
the precision trap of $10~\mbox{s}$ has been employed prior to the
frequency determination.}}}
\end{figure}

{\bf $^{186}\mbox{Tl}^{m}$}: An unambiguous assignment of the
ISOLTRAP mass value to one of the long-lived states in
$^{186}\mbox{Tl}$ is not straightforward. This is also reflected
by the different spin assignments given in NUBASE (Tab.
\ref{A_kap5_NUBASE}) and in the Evaluated Nuclear Structure Data
Files (ENSDF). Mass values of both excited states are given with
uncertainties of $50~\mbox{keV}$. The energy difference between
$^{186}\mbox{Tl}^{m}$ and $^{186}\mbox{Tl}^{n}$ is
$373.9(5)~\mbox{keV}$ \cite{VanD1991}. The attempt to employ a
very long RF-excitation of $9~\mbox{s}$, corresponding to a mass
resolution of $\Delta m = 33~\mbox{keV}$, failed. For excitation
times beyond $3~\mbox{s}$, a decay of the stored ions has been
observed (see half-lives in Tab. \ref{A_kap5_NUBASE}).\\
Figure \ref{Tl186} shows the results of the four cyclotron
frequency measurements. With a mass resolution of $210~\mbox{keV}$
(measurement \# 2 - 4) a mixture between $^{186}\mbox{Tl}^{m}$ and
$^{186}\mbox{Tl}^{n}$ is resolved and the presence of either of
the states should have been detected. Measurement \# 2 and \# 4
yield count-rate-class corrected frequencies. The slope $z$ in
[mHz/ion] of the linear extrapolation of the cyclotron frequency
of $^{186}\mbox{Tl}^+$ is indicated for each measurement. The
third data point (\# 3), which has been obtained after an
additional waiting time of $10~\mbox{s}$ in the upper precision
trap, does not show a significant count rate effect, since this
measurement allowed for a decay of the $T_{1/2} =
2.9~\mbox{s}$-state. To avoid any undetected contamination from
the low-resolution measurement \# 1, it was not considered in the
final weighted average of the frequency ratio. After a comparison
to the spin systematics, as given in Tab. \ref{A_kap5_NUBASE}, and
the agreement of the ISOLTRAP value with the intermediate
evaluation IAME2002 this result is assigned to the
$^{186}\mbox{Tl}^{m}$ state. It is not excluded that a small
contamination by the ground
state ($\Delta E = 40~\mbox{keV}$) might be present.\\

{\bf $^{196}\mbox{Tl}^{m}$}: The mass resolution, being
$120~\mbox{keV}$ during this measurement, was sufficiently high to
separate both states of $^{196}\mbox{Tl}$. The resulting ISOLTRAP
value agrees with the mass value of the excited, medium spin state
$^{196}\mbox{Tl}^{m}$ ($T_{1/2}=1.41~\mbox{h}$, $I^\pi = (7^+)$)
though the uncertainties of the IAME2002 values are rather
large.\\

{\bf $^{190,194,196}\mbox{Bi}$}: These nuclides have three
long-living states similar to almost all even-$A$ bismuth isotopes
up to {$^{212}\mbox{Bi}$}. The excitation energies of about
$150~\mbox{keV}$ are all known with uncertainties of around
$70~\mbox{keV}$, except for $^{196}\mbox{Bi}$ (Fig. \ref{even}).
The resolving power employed during those cyclotron frequency
measurements of approximately $3~\times~10^{5}$ corresponds to a
mass resolution between $560$ and $590~\mbox{keV}$. In order to
obtain a mass resolution of $50~\mbox{keV}$ an excitation time of
$7~\mbox{s}$ is required.\\
The mass value of ISOLTRAP for $^{196}\mbox{Bi}$ agrees well to
the one of the $0.6~\mbox{s}$-state ($^{196}\mbox{Bi}^{m}$).
However, a contamination cannot be excluded. Accounting for a
possible admixture of either the $3^+$ or the $10^-$ state the
original uncertainty of $14~\mbox{keV}$ had been increased to
$50~\mbox{keV}$ in the AME2003.\\
Due to the good agreement for $^{196}\mbox{Bi}$, it is concluded
that the medium spin state has been predominantly produced for
$^{194}\mbox{Bi}$ as well. Since the possible presence of both
isomeric states, $3^+$ and $10^-$, whereas each of them is about
$100~\mbox{keV}$ different in energy, cannot be excluded with a
resolution of $580~\mbox{keV}$, the original error of
$14~\mbox{keV}$ is increased in the Atomic Mass
Evaluation 2003 to $50~\mbox{keV}$.\\
In $^{190}\mbox{Bi}$ the medium
spin state exhibits a half-life in the nanoseconds regime.\\
The measured ISOLTRAP value with $25~\mbox{keV}$ uncertainty
agrees within $1~\sigma$ to the value of the $6.2~\mbox{s}$-state
($^{190}\mbox{Bi}$). Due to the minor resolution during the
measurement, a mixture of both states $^{190}\mbox{Bi}$ and
$^{190}\mbox{Bi}^{m}$ differing by $E = 150(190)~\mbox{keV}$
might be present.\\

{\it {\bf $^{192}\mbox{Bi}$}}: In contrast to other even-$A$
bismuth isotopes from $^{190}\mbox{Bi}$ to $^{210}\mbox{Bi}$, no
medium nuclear spin of, for example, $I = 6$ or $I = 7$ has been
observed. However, the general trend in this experiment points to
a favored production of medium spin states, $I = 9/2$ or 7. The
question on the production ratio between the $3^+$ ground state
and the excited $10^-$ state remains unsolved. From the general
trend throughout all these measurements it is observed, that it is
the higher spin state that is produced with a higher weight and
measured in the trap. Consequently the mass value has been
tentatively assigned to $^{192}\mbox{Bi}^{m}$ in the Atomic Mass
Evaluation 2003. In the mass measurement of this nuclide two
observations are made: First, the counting class analysis of two
individual measurements with a resolution of $\Delta m =
110~\mbox{keV}$ indicate the presence of a contamination and give
a shifted, extrapolated value. Secondly, the characteristic
pattern for an in-trap decay with short half-life is observed in
the time-of-flight spectrum although the half-lives of both known
states are more than ten times longer than the excitation time of
$T_{\mbox{\scriptsize{RF}}} = 3~\mbox{s}$. This might indicate a
third, so far unknown, isomeric state with shorter half-life or
any other contamination.
\subsection{Isobaric nuclides {\bf $^{229}\mbox{Fr}$} and {\bf $^{229}\mbox{Ra}$}}
\begin{figure}[h]
\centerline{\mbox{\includegraphics[width=0.7\textwidth]{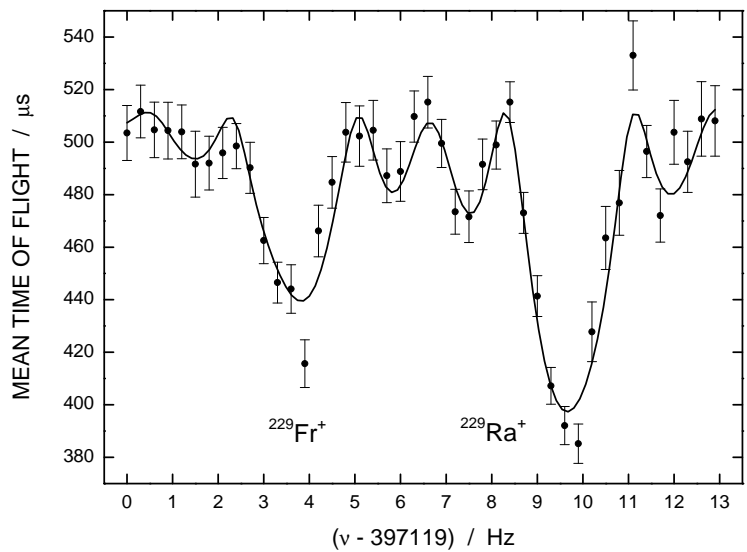}}}
\centerline{\parbox{0.75\textwidth}{\caption{\label{Fr229_Ra229}
Cyclotron resonance curves of the isobars $^{229}\mbox{Fr}$ (left)
and $^{229}\mbox{Ra}$ (right). Both isobars were simultaneously
delivered from the preparation trap to the precision trap.}}}
\end{figure}

\begin{figure}
\centerline{\mbox{\includegraphics[width=0.89\textwidth]{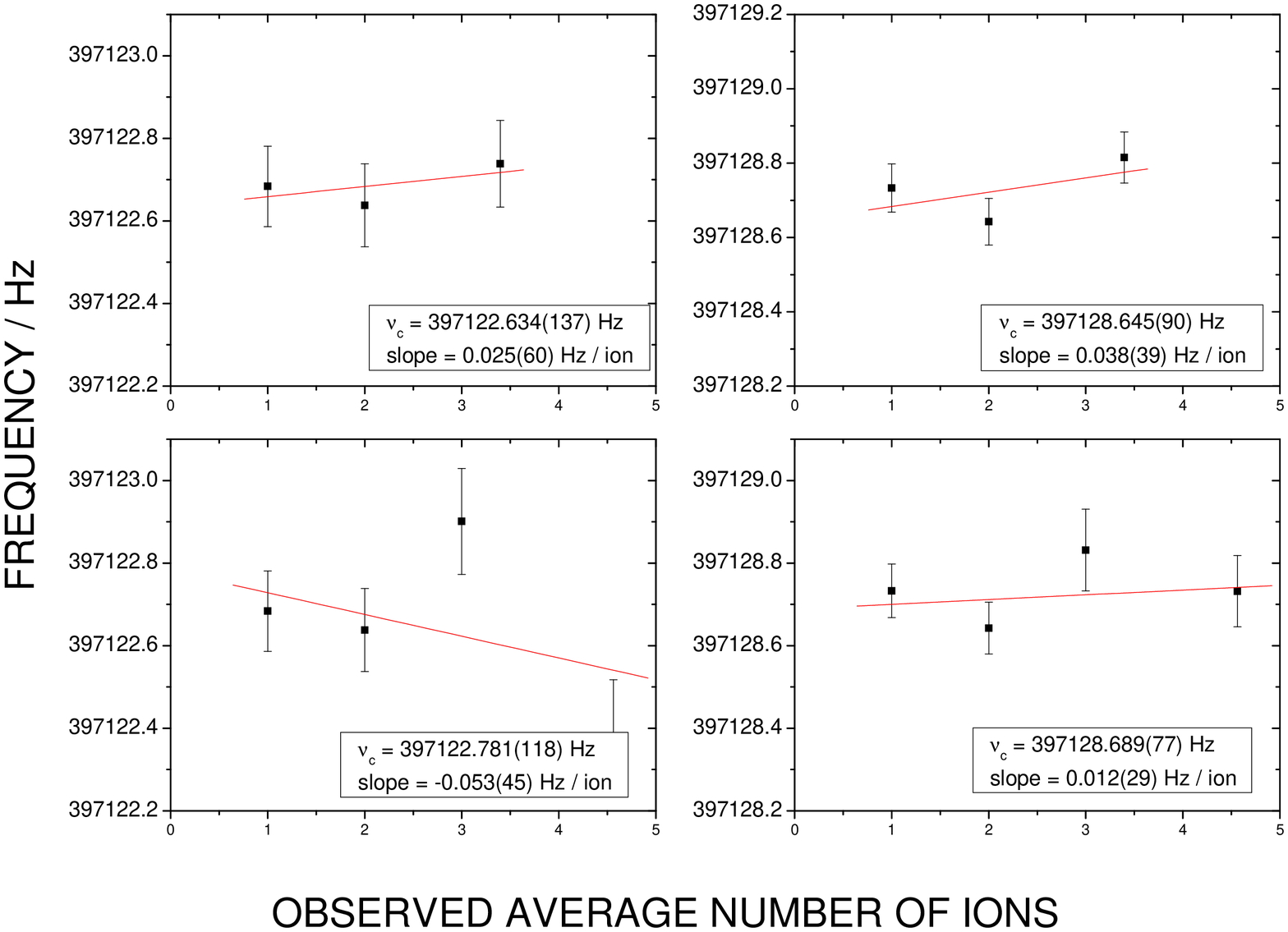}}}
\centerline{\parbox{0.75\textwidth}{\caption{\label{z-class_229}Count-rate-class
analysis for isobaric ions $^{229}\mbox{Fr}$ (left) and
$^{229}\mbox{Ra}$ (right) of the signal shown in Fig.
\ref{Fr229_Ra229}. All registered ions were subdivided into three
(upper diagrams) as well as into four (lower diagrams) count rate
classes. The data points are fitted by linear regression. The
resulting slopes and center frequencies of the cyclotron
resonances are given in the inset. None of the individual
measurements shows a significant variation of the cyclotron
frequency on the number of detected ions.}}}
\end{figure}

The nuclides $^{229}\mbox{Fr}$ and $^{229}\mbox{Ra}$ were stored
simultaneously in the precision trap since the applied mass
resolving power $R \approx 4 \times 10^{4}$ in the preparation
Penning trap was not sufficient to separate the $A=229$-isobars.
Figure \ref{Fr229_Ra229} shows the cyclotron resonance curves of
the isobars, which differ in mass by $3.2~\mbox{MeV}$.
The center frequencies including the first sidebands are clearly
separated with a resolving power of $R \approx 2.4 \times 10^{5}$,
using an excitation time of $T_{\mbox{\scriptsize{RF}}} =
600~\mbox{ms}$. A superposition of two individual cyclotron
resonance curves, weighed by their relative abundance, was fitted
to the data points. Assuming a detection efficiency of the
microchannel plate of 30 \%, the average number of detected ions
of 2.5 corresponds to about eight ions that were simultaneously
stored. Therefore, the possibility of a frequency shift dependent
on the number of ions (see Sec. \ref{counti}) was investigated
in a detailed count-rate-class analysis.\\
The result of such an analysis is plotted in Fig.
\ref{z-class_229} for the cyclotron frequencies of
$^{229}\mbox{Fr}$ (left) and $^{229}\mbox{Ra}$ (right). Here, a
total number of 2340 registered ions was subdivided into three
(upper diagrams) as well as into four (lower diagrams) count-rate
classes. {\it Different} subdivisions were made in order to
exclude any possible count-rate dependance produced within the
statistics. In both cases no significant count-rate dependance is
observed.
\subsection{Other nuclides}
\begin{figure}
\centerline{\mbox{\includegraphics[width=0.7\textwidth]{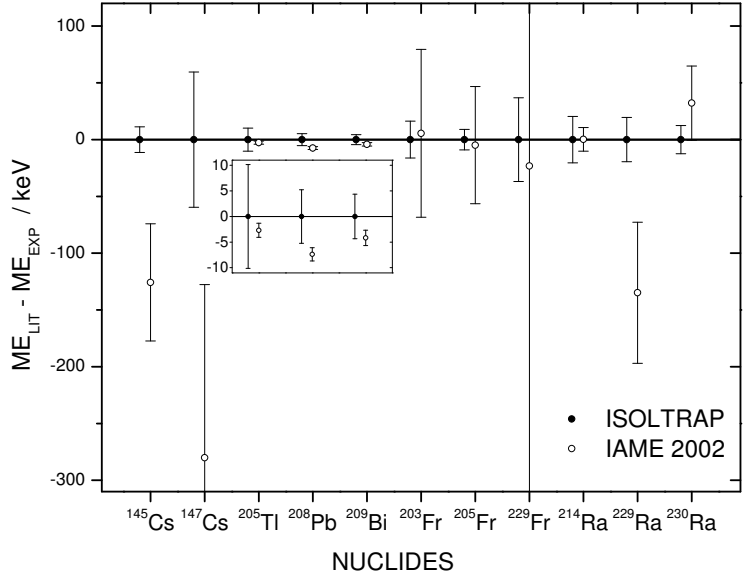}}}
\centerline{\parbox{0.8\textwidth}{\caption{\label{rest}Comparison
of ISOLTRAP mass values to the intermediate adjustment of the
Atomic Mass Evaluation IAME2002. The line at zero represents the
experimental ISOLTRAP values with their uncertainties. The inset
shows an enlarged scale for the stable nuclides.}}}
\end{figure}

Figure \ref{rest} shows the mass excess values of other nuclides
studied in this work, including the stable isotopes
$^{205}\mbox{Tl}$, $^{208}\mbox{Pb}$, and $^{209}\mbox{Bi}$. The
masses of these stable isotopes, which are known with high
accuracy, agree within one standard deviation with IAME2002. The
largest discrepancies are observed for $^{145}\mbox{Cs}$
($-2.4~\sigma$), $^{147}\mbox{Cs}$ ($-1.7~\sigma$), and
$^{229}\mbox{Ra}$ ($-2.0~\sigma$).
\clearpage
\subsection{\label{ESRcomp}Comparison to other experimental data}
\begin{figure}
\begin{center}
\mbox{\includegraphics[width=0.75\textwidth]{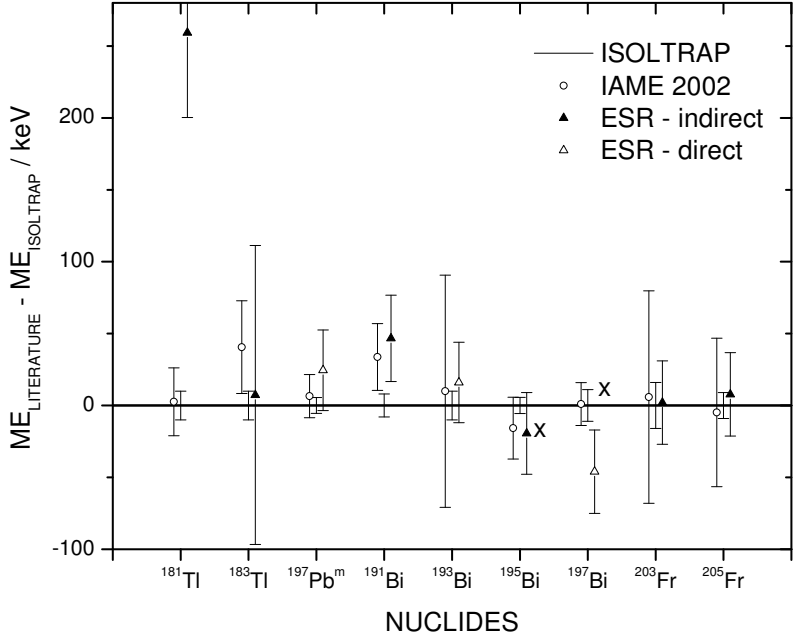}}
\end{center}
\centerline{\parbox{0.75\textwidth}{\caption{\label{TRAP-ESR}Comparison
of ISOLTRAP, ESR data, and the values of the intermediate
adjustment IAME2002 \cite{Audi2003a}. The ESR values are
distinguished between direct measurements (open triangles) and
indirect (closed triangles) data which are linked to direct ESR
measurements by $Q$ values from literature. All directly
determined ESR values are taken from \cite{Litv2005}, whereas the
indirectly determined ones are from \cite{Litv2003}. For some of
the latter cases a direct value of a mixture of isomeric states is
given in \cite{Litv2005}. The symbol 'x' denotes those nuclides
where an isomeric mixture has not been resolved.}}}
\end{figure}

Some of the mass values which are reported in this work have also
been determined by Schottky mass spectrometry in the Experimental
Storage Ring ESR at GSI \cite{Litv2003,Litv2005}. Figure
\ref{TRAP-ESR} compares the results of both experiments and the
values of the IAME2002. A larger set of data is obtained by
including also the indirectly determined results at ESR using $Q$
values from literature \cite{Litv2003}. In most of the cases the
data from both measurement techniques agree within their
uncertainties. An exception is the nuclide $^{181}\mbox{Tl}$,
where both values differ by more than $4.3~\sigma$. Since the ESR
value has been indirectly determined, this discrepancy could be
attributed as well to the $Q_\alpha$ value.
\section{\label{AMERES}Results}
Table \ref{A_kap5_MA8_input} compiles the mass excesses of those
radionuclides discussed here and compares the data with values
given in NUBASE 1997 \cite{Audi1997} and \cite{Audi2003a}. In
addition, the results of the atomic mass evaluation published in
2003 (AME2003) are given, which include the ISOLTRAP results. Due
to the manifold interconnections, the present ISOLTRAP data
improve not only the directly measured nuclides, but also a large
region from $Z = 73$ to $Z = 92$ interconnected via $\alpha$
chains in particular. Fig. \ref{linked nuclides} shows the
respective section of the nuclear chart from tantalum to neptunium
isotopes, indicating all related nuclides in this region.\\
For primary nuclides, which directly contribute to the adjustment
procedure, the total influence of each input value in the AME is
given in Tab. \ref{A_kap5_MA8_input} by the {\it significance} in
per cent. It is the sum of all influences on individual mass
values $m_{\lambda}$. The quantity 'v/s' denotes the difference
between both values, divided by the uncertainty of the ISOLTRAP
value. The last column marks those data which were not used in the
adjustment procedure. This is the case, if, for example, a
contamination of the ISOLTRAP measurement by a further isomeric
state could not be excluded, or if, for the stable nuclides,
another high-accuracy mass determination already exists, which is
at least three times more precise than the ISOLTRAP value.\\
In addition, the mass values of around 100 other nuclides changed
after the atomic mass evaluation with the ISOLTRAP data reported
here. The largest modifications in the adjusted mass values is
observed for $^{145}\mbox{Ba}$ ($7~\sigma$).\\
\begin{figure}
\begin{center}
\mbox{\includegraphics[width=0.97\textwidth]{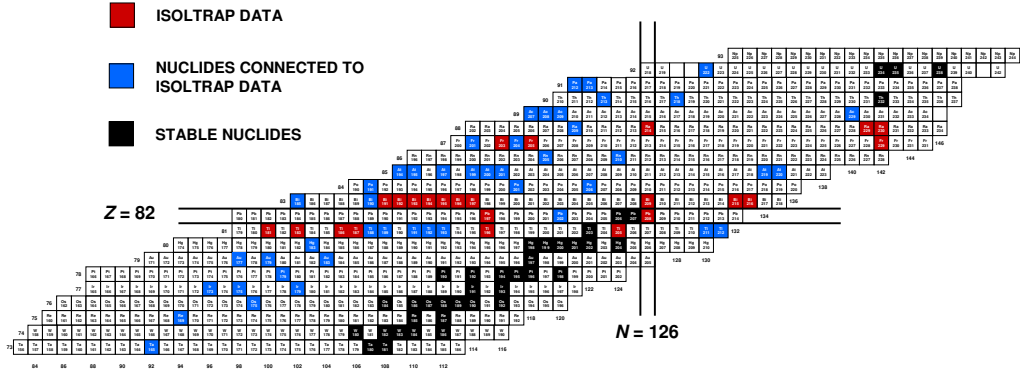}}
\end{center}
\centerline{\parbox{0.8\textwidth}{\caption{\label{linked
nuclides}Region of the nuclear chart around $Z = 82$ and $N =
126$, indicating the nuclides measured by ISOLTRAP (in red). All
other nuclides which are connected to ISOLTRAP data are shown (in
blue). Color figure online.}}}
\end{figure}
\begin{center}
\begin{longtable}{lrrrcrc}
\caption{\label{A_kap5_MA8_input}ISOLTRAP mass data compared to
the data given in NUBASE \cite{Audi1997} and \cite{Audi2003a}, and
the Atomic Mass Evaluation AME2003 \cite{Audi2003b}. Here, mass
excess values are rounded as described at p. 344 in
\cite{Audi2003b}. The significance of the ISOLTRAP measurement is
listed in column 'Sign.'. Column 'v/s' denotes the deviation
between the mass value as determined by ISOLTRAP and the AME2003
divided by the uncertainty in the input value. The following
abbreviations are used in the column 'Remark': 'U' denotes
nuclides, that have not been used in the adjustment procedure due
to their low impact. 'F' indicates nuclides with a possible
presence of a contamination which, for this reason, are excluded
from the adjustment. 'E': The ISOLTRAP value determines the
excited $^{196}\mbox{Tl}^{m}$ state with an excitation energy of
$E = 394.2(5)~\mbox{keV}$, whereas the mass for the ground state
has been implemented in the AME.}
\\\hline
Nuclide  &  \multicolumn{3}{c}{Mass excess [keV]}  &  Sign. & v/s & Remark \\
         &   {ISOLTRAP} &  {NUBASE1997} and \cite{Audi2003a} &  {AME2003} & \% & & \\
         \endfirsthead
\caption{(Continued).}
\\\hline
Nuclide  &  \multicolumn{3}{c}{Mass excess [keV]}  &  Sign. & v/s & Remark \\
         &   {ISOLTRAP} & {NUBASE1997} and \cite{Audi2003a} &  {AME2003} & \% & & \\
\hline
\endhead
\hline
$^{145}\mbox{Cs}$       &  -60052(11)   &   -60180(50)     &   -60057(11)    &   94  &   -0.4    &       \\
$^{147}\mbox{Cs}$       &  -52011(60)   &   -52290(150)    &   -52020(50)    &   79  &   -0.1    &       \\
$^{181}\mbox{Tl}$       &  -12802(10)   &   -12490\#(300\#)&   -12801(9)   &   92  &   -0.2     &       \\
$^{183}\mbox{Tl}$       &  -16592(10)   &   -16290\#(190\#)&   -16587(10)    &   91  &   0.4     &       \\
$^{186}\mbox{Tl}^{m}$   &  -19874.4(8.6)&   -19980(50)     &   -19874(9)   &   100 &           &       \\
$^{187}\mbox{Tl}^{m}$   &  -22154(23)   &   -22085(22)     &   -22109(8)   &       &   2       &   F   \\
$^{196}\mbox{Tl}^{(m)}$ &  -27103(12)   &   -27470\#(100\#)&   -27497(12)    &   100 &           &   E   \\
$^{205}\mbox{Tl}$       &  -23818(10)   &   -23820.8(1.4)  &   -23820.6(1.3)   &       &   -0.3    &   U   \\
$^{197}\mbox{Pb}^{m}$   &  -24429.5(5.5)&   -24423(15)     &   -24429(6)   &   100 &           &       \\
$^{208}\mbox{Pb}$       &  -21742.2(5.2)&   -21749.6(1.3)  &   -21748.5(1.2) &       &   -1.2    &   U   \\
$^{190}\mbox{Bi}^{m}$   &  -10535(25)   &   -10590(50)     &   -10483(10) &       &   2.1     &   F   \\
$^{191}\mbox{Bi}$       &  -13244.9(8.0)&   -13211(23)     &   -13240(7) &   86  &   0.6     &       \\
$^{192}\mbox{Bi}^{m}$   &  -13398.5(9.0)&   -13220\#(120\#)&   -13399(9) &   100 &           & \footnote{Tentative assignment}\\
$^{193}\mbox{Bi}$       &  -15875(10)   &   -15870(80)     &   -15873(10) &   100 &   0.2     &       \\
$^{194}\mbox{Bi}^{m}$   &  -15878(50)   &   -15970\#(440\#)&   -15880(50)    &   100 &           &       \\
$^{195}\mbox{Bi}$       &  -18023.7(5.6)&   -18039(21)     &   -18024(6) &   100 &           &       \\
$^{196}\mbox{Bi}^{m}$   &  -17868(50)   &   -17856(30)     &   -17842(25)    &   100 &   0.5     &       \\
$^{197}\mbox{Bi}$       &  -19706(25)   &   -19689(15)     &   -19688(8) &   100 &   0.8     &       \\
$^{209}\mbox{Bi}$       &  -18254.9(4.3)&   -18259.1(1.5)  &   -18258.5(1.4) &       &   -0.8    &   U   \\
$^{215}\mbox{Bi}$       &  1648(15)     &   1710(100)      &   1649(15)      &   100 &           &       \\
$^{216}\mbox{Bi}$       &  5873(11)     &   5780\#(100\#)  &   5874(11)      &   100 &           &       \\
$^{203}\mbox{Fr}$       &  861(16)      &   870(70)        &   861(16)       &   100 &           &       \\
$^{205}\mbox{Fr}$       &  -1308.6(9.0) &   -1310(50)      &  -1310(8)   &   100 &   -0.1    &       \\
$^{229}\mbox{Fr}$       &  35816(37)    &   35790\#(360\#) &   35820(40)     &   100 &           &       \\
$^{214}\mbox{Ra}$       &  100(20)      &   100(10)        &   101(9)    &   100 &   0       &       \\
$^{229}\mbox{Ra}$       &  32575(19)    &   32440(60)      &   32563(19)     &   91  &   -0.6    &       \\
$^{230}\mbox{Ra}$       &  34518(12)    &   34550(30)      &   34518(12)     &   100 &           &       \\
\hline
\end{longtable}
\end{center}
\section{Discussion}
\subsection{Irregularities of the mass surface}\label{nucstruc}
\begin{figure}
\centerline{\mbox{\includegraphics[width=0.89\textwidth]{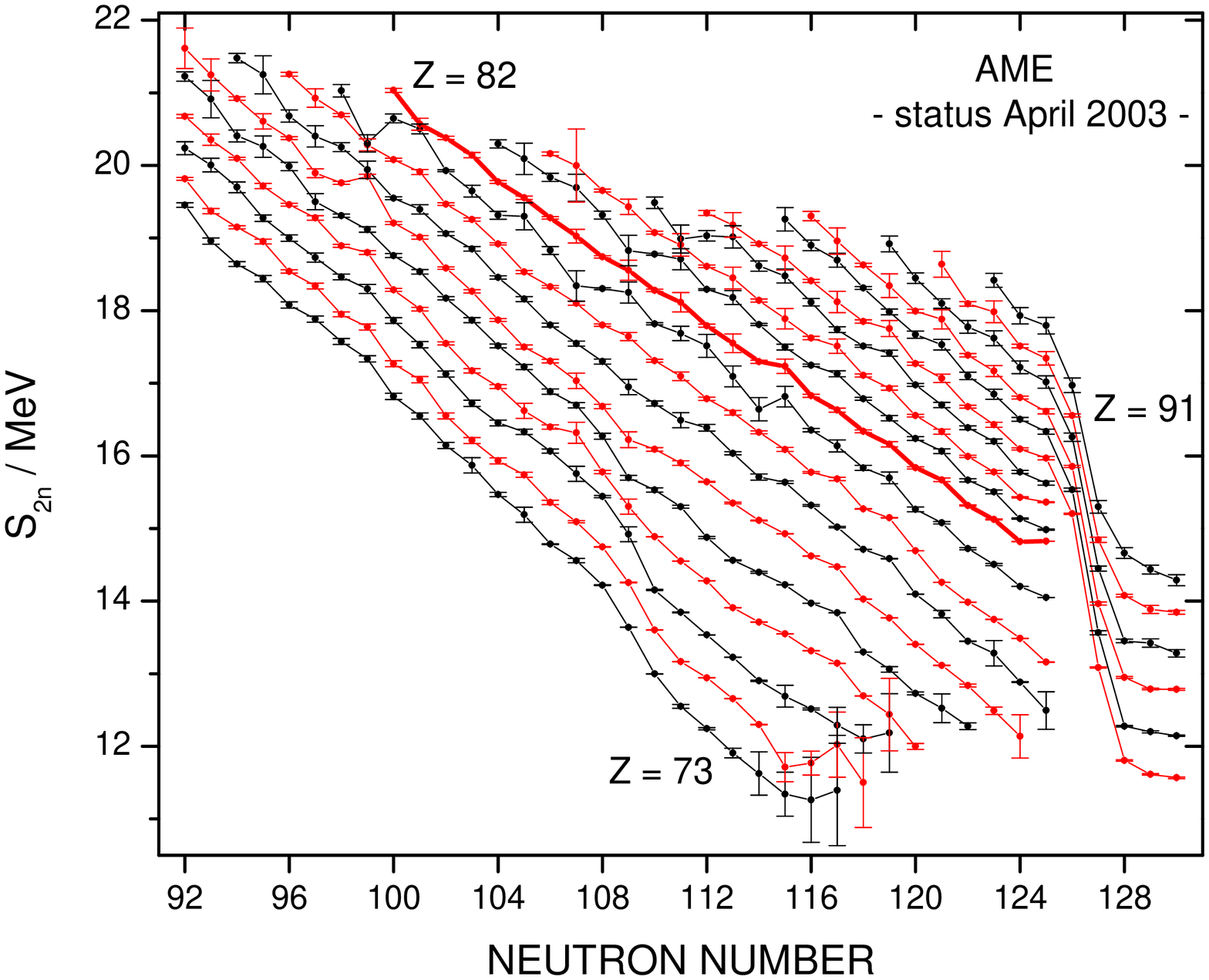}}}
\centerline{\parbox{0.85\textwidth}{\caption{\label{bbb}Comparison
of two-neutron separation energies $S_{\mbox{\scriptsize 2n}}$ as
obtained in AME1995 \cite{Audi1995} and 2003. Color figure
online.}}}
\end{figure}

\begin{figure}
\centerline{\mbox{\includegraphics[width=0.85\textwidth]{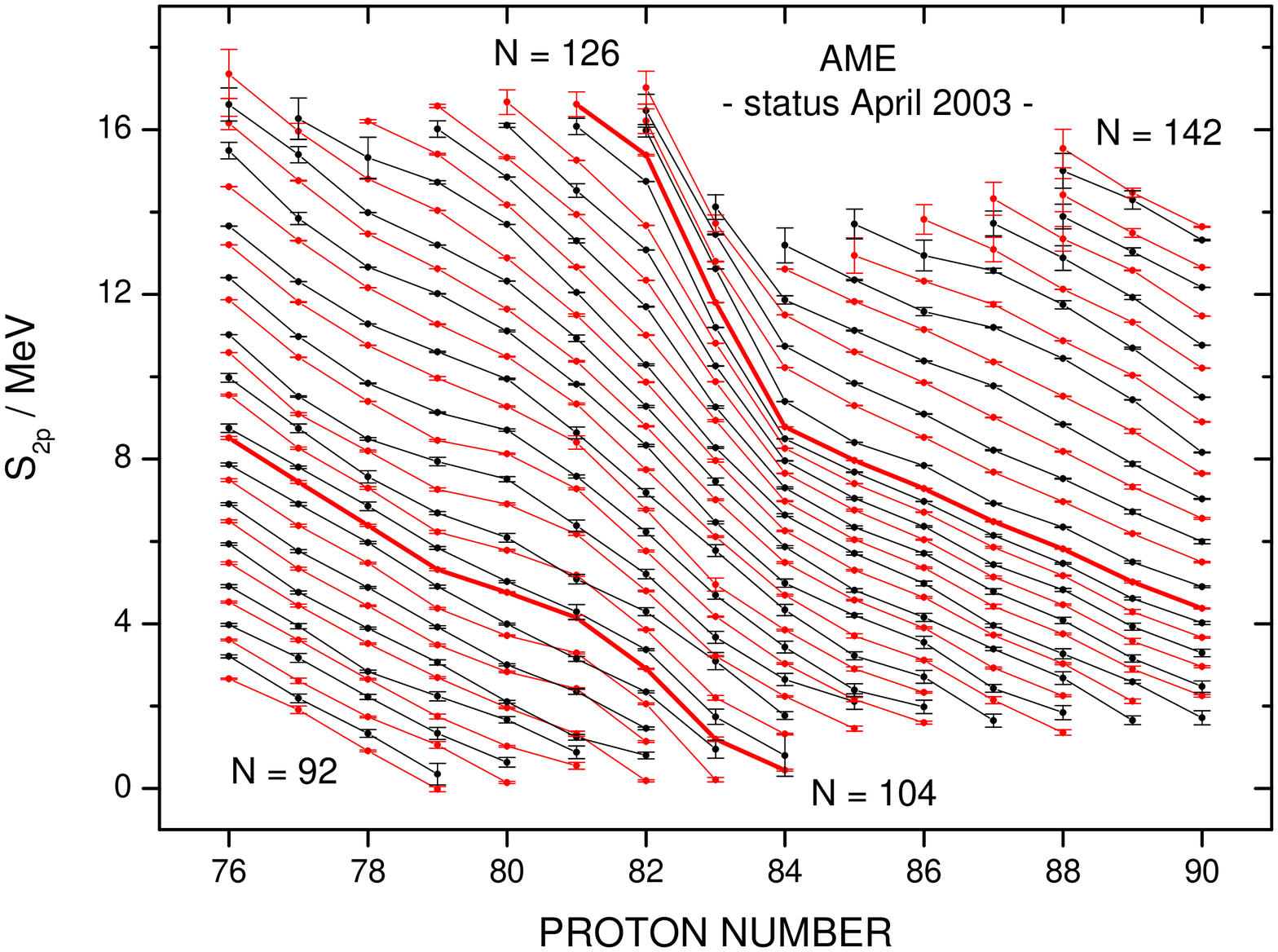}}}
\centerline{\parbox{0.85\textwidth}{\caption{\label{ddd}Comparison
of two-proton separation energies $S_{\mbox{\scriptsize 2p}}$ as
obtained in AME1995 \cite{Audi1995} and 2003. Color figure
online.}}}
\end{figure}

The mass values and the corresponding nuclear binding energies
represent the underlying forces acting in a nucleus. In analogy to
the ionization potential of atoms, plots of separation energies
for the removal of two protons,
\begin{equation}
    \label{S2p}
S_{\mbox{\scriptsize 2p}} = BE (Z,N) - BE (Z-2,N),
\end{equation}
or two neutrons,
\begin{equation}
    \label{S2n}
S_{\mbox{\scriptsize 2n}} = BE (Z,N) - BE (Z,N-2),
\end{equation}
reveal the details of nuclear shell structure. In
this representation, odd-even effects caused by the usually much
larger binding energy for even proton or neutron
numbers are removed.\\
Figures \ref{bbb} and \ref{ddd} show the behavior of separation
energies for protons and neutrons versus the respective nucleon
number. The results from AME1995 and AME2003 are compared. A
global trend, a continuous decrease in the separation energies, is
observed. Discontinuities from this general trend appear at shell
closures or in regions of strong deformations. The $N = 126$ shell
closure is clearly visible in Fig. \ref{bbb}. The much less
pronounced strength of the proton shell closure $Z=82$ is larger
at the magic neutron number $N =126$ than at the half-filled shell
$N = 104$ (Fig. \ref{ddd}). The main differences between the
appearance of the mass surface in 1995 and 2003 are substantially
reduced uncertainties, due to improved as well as numerous new
experimental input data. These are ISOLTRAP measurements in the
isotopic chains of mercury, platinum, lead, polonium, radon, and
radium \cite{Schw1998,Schw2001,Kohl1999} as well as a large number
of new results obtained at the ESR by the Schottky technique
\cite{Rado2000,Litv2003,Litv2005}. In particular the small
uncertainty ($\delta m \approx 25~\mbox{keV}$) of data from
ISOLTRAP along the extended isotopic chain from $^{179}\mbox{Hg}$
to $^{195}\mbox{Hg}$ enables the observation of finer nuclear
structure effects and to study, for example, nuclear shape
coexistence and the influence of low-excitation intruder states
\cite{Schw2001,Foss2002,Juli2001}. Note the pronounced
irregularities in the
thallium isotopic chain at neutron numbers $N = 99, 105, 107,$ and 114.\\
The two-proton separation energies are much more regular and
smooth in this region of the nuclear chart than those for two
neutrons. However, a peculiar irregularity in the two-proton
separation energies appears for the isotopes of thallium ($Z =
81$) with even neutron numbers $N$ from 98 to 108: Some separation
energies are nearly equal or even higher than those of the
isotopes with one additional neutron. Such an effect is very
uncommon in the chart of nuclei. \clearpage
\begin{figure}
\centerline{\mbox{\includegraphics[width=0.85\textwidth]{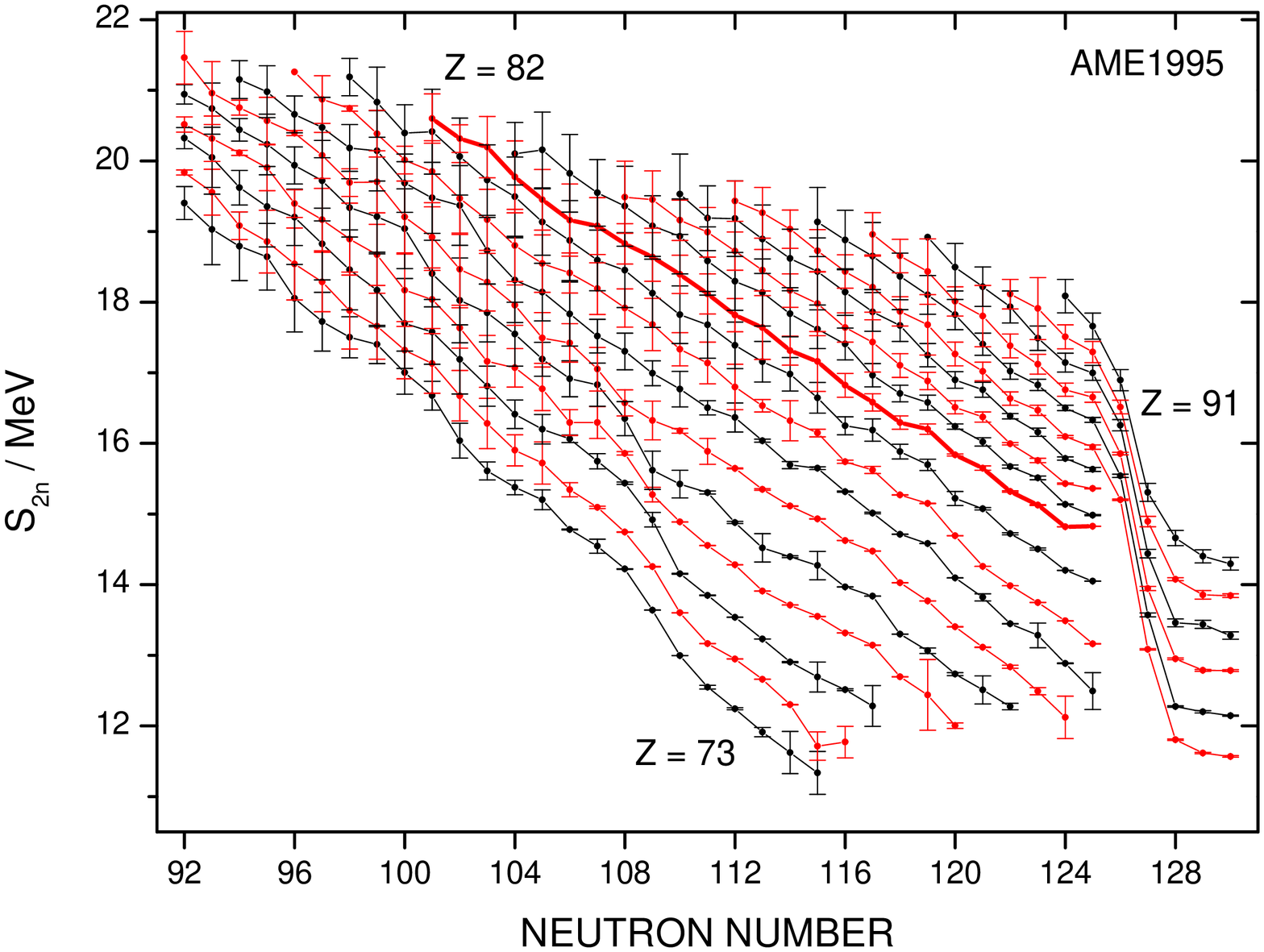}}}
\end{figure}
\clearpage
\begin{figure}
\centerline{\mbox{\includegraphics[width=0.85\textwidth]{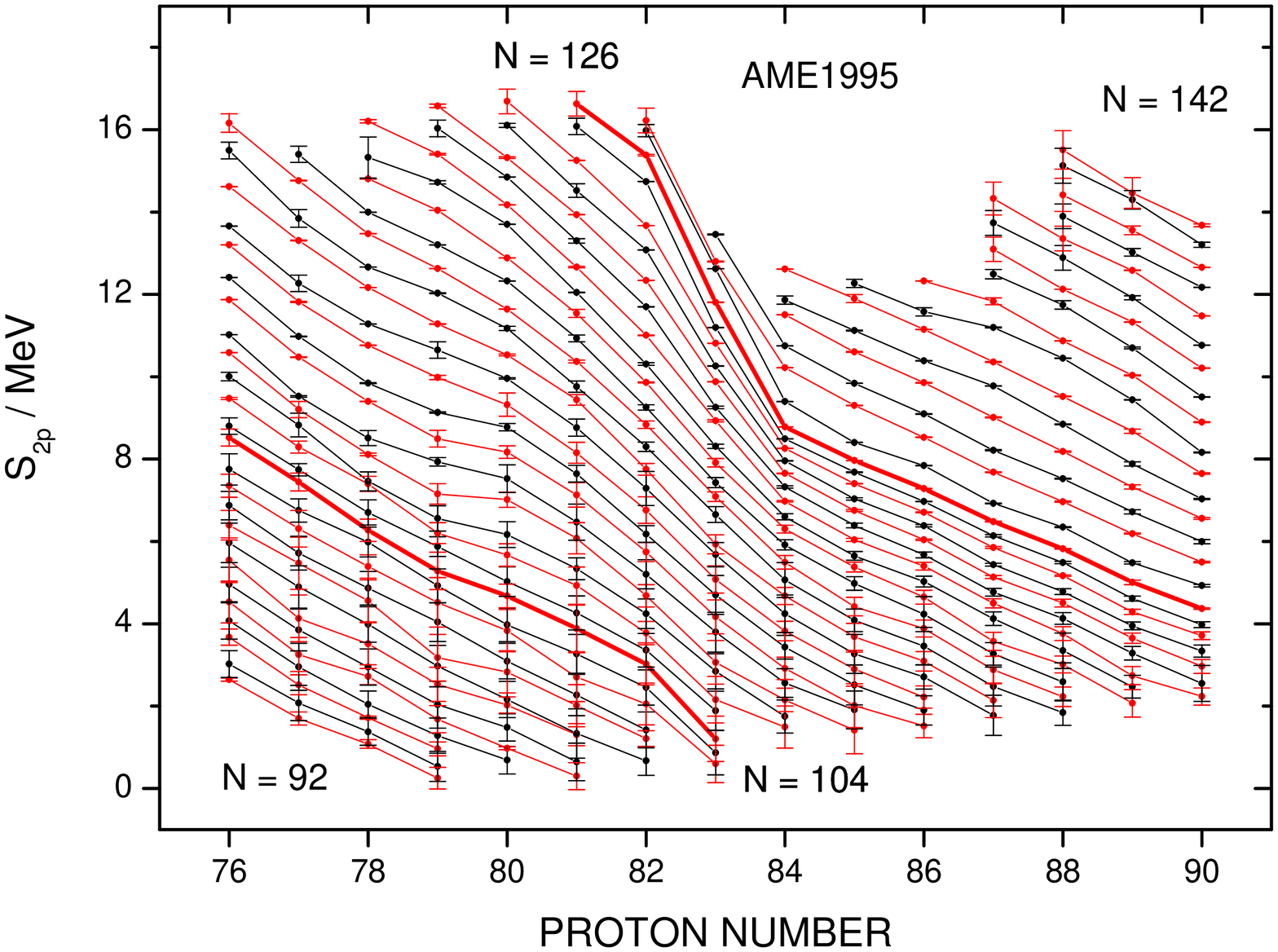}}}
\end{figure}
\clearpage

\begin{figure}
\centerline{\mbox{\includegraphics[width=0.95\textwidth]{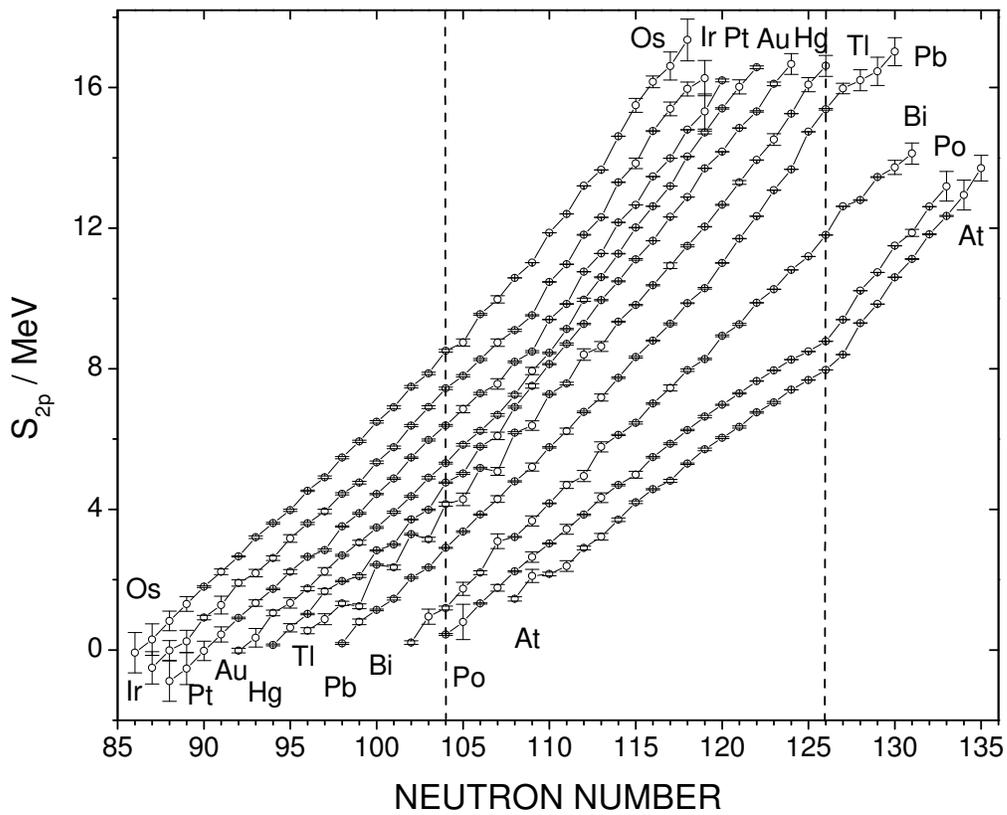}}}
\centerline{\parbox{0.75\textwidth}{\caption{\label{stagg}Two-proton
separation energies as a function of neutron number.}}}
\end{figure}

In order to investigate this anomalous behavior further, the
two-proton separation energies are plotted in Fig. \ref{stagg}
against the neutron number and the isotopes of an element are
connected by lines. This presentation shows very clearly in the
region of the lead isotopes at $Z = 82$ the steadily increasing
shell gap from mid-shell $N = 104$ towards the magic neutron
number $N = 126$. Also the above mentioned irregularities in the
thallium and (less pronounced) in the mercury isotopes are
revealed distinctly as an odd-even effect. The reason for this
effect is unclear. Three possible explanations
are discussed in the following:\\
1. The masses of $^{181}\mbox{Tl}$ and $^{183}\mbox{Tl}$,
determined with ISOLTRAP cyclotron resonances and included in
AME2003, were wrongly assigned to the ground states of these
isotopes. However, an assignment to the isomeric states
(excitation energy $E \ge 800~\mbox{keV}$ or $E \ge
600~\mbox{keV}$) would increase the staggering even further. Also
the mass of $^{187}\mbox{Tl}$ is determined to more than 60 \% by
ISOLTRAP, but in this case indirectly via the $Q$ value of the
$\alpha$ decay and the directly measured mass of
$^{191}\mbox{Bi}$. Also here, a possible problem in the assignment
can be excluded due to the sufficiently high resolving power for
separating ground and isomeric states and
because of the short half-life of the isomer.\\
2. Some mass measurements might be wrong. This explanation is
contradictory to the fact that the mass values in question stem
from different sources and mass spectrometric techniques
(ISOLTRAP, ESR, $Q$ values, etc.). Furthermore, the odd-even
staggering is not an isolated phenomenon but appears also (less
pronounced) in the neighboring isotopic chain of mercury. This is
not easily explicable since the mercury and the thallium isotopes
are not interconnected by $Q_\alpha$ values.\\
3. The odd-even staggering of the two-proton separation energies
in the thallium isotopic chain is factual and represents a nuclear
structure effect. This would imply a stronger pairing effect for
neutron numbers around mid-shell ($N = 104$) than for nearly
double-magic thallium nuclei with $Z = 81$ and $N \approx 126$.
Similar to the drastic odd-even staggering observed in the charge
radii of the mercury isotopes (caused by shape transitions)
\cite{Ulm1986} these phenomena depend very critically on the
neutron and proton configurations and very elaborated calculations
are required to reproduce these effects.
\clearpage
\subsection{Correlation between pairing energies and nuclear charge radii}
\label{sec:2}
\begin{figure}
\begin{center}
\resizebox{0.80\textwidth}{!}{%
  \includegraphics{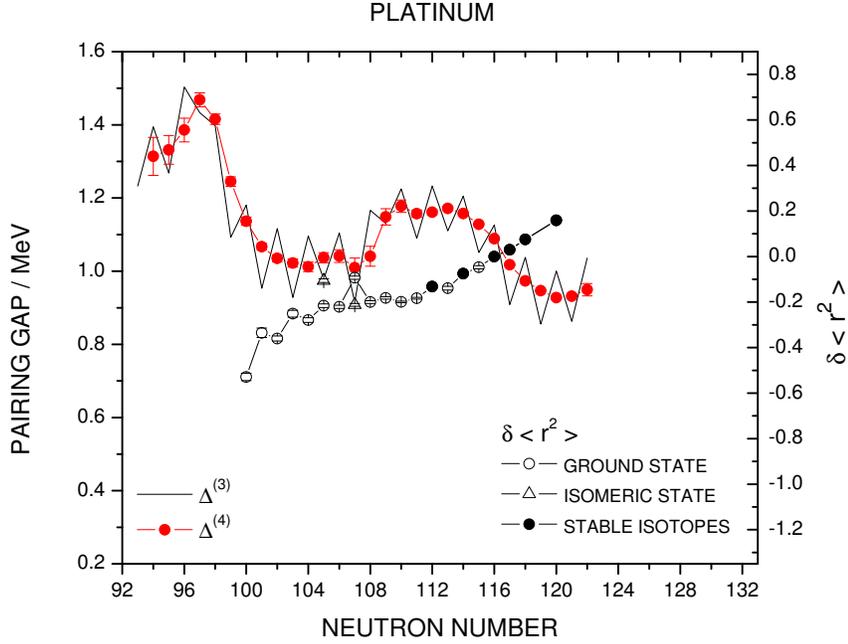}}
\end{center}
\caption{Comparison of neutron pairing gap energies $\Delta^{(3)}$
and $\Delta^{(4)}$ with nuclear mean square charge radii
$\delta<r^2>$ for the isotopic chain of platinum. The
uncertainties of the $\delta<r^2>$ values are smaller than the
size of the symbols. Color figure online.} \label{figrad1}
\end{figure}

\begin{figure}
\begin{center}
\resizebox{0.80\textwidth}{!}{%
  \includegraphics{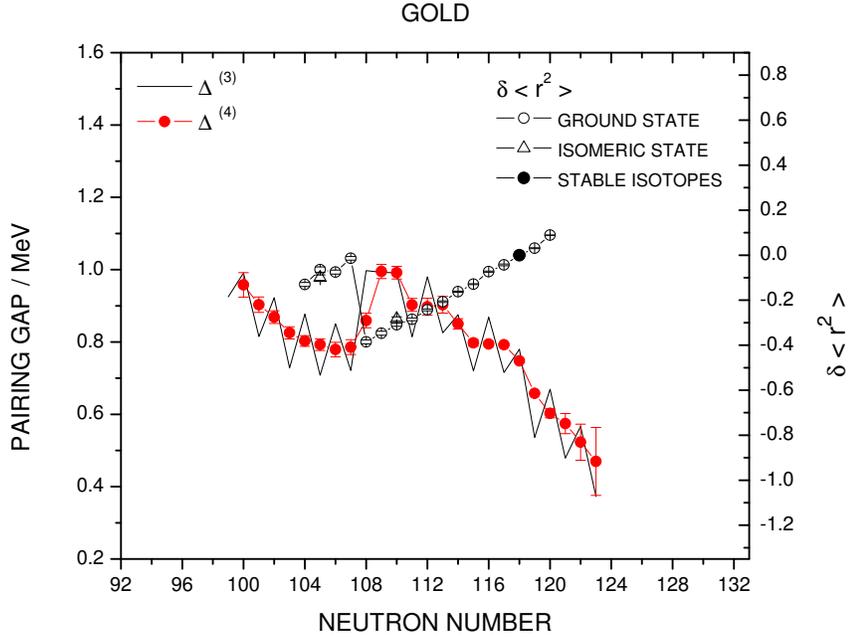}}
\end{center}
\caption{Same as Fig. \ref{figrad1} but for gold isotopes.}
\label{figrad2}
\end{figure}

\begin{figure}
\begin{center}
\resizebox{0.80\textwidth}{!}{%
  \includegraphics{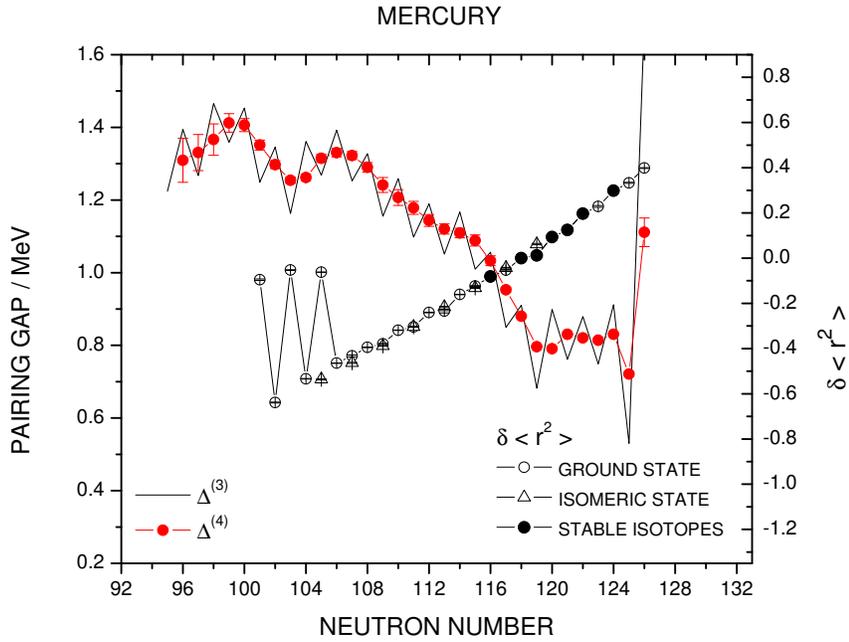}}
\end{center}
\caption{Same as Fig. \ref{figrad1} but for mercury isotopes.}
\label{figrad3}
\end{figure}

\begin{figure}
\begin{center}
\resizebox{0.80\textwidth}{!}{%
  \includegraphics{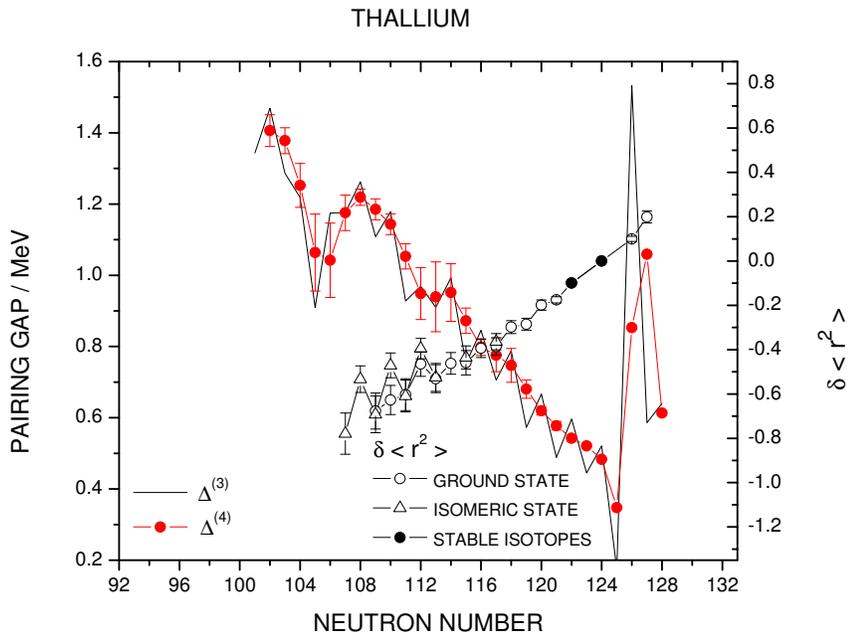}}
\end{center}
\caption{Same as Fig. \ref{figrad1} but for thallium isotopes.
Note that the radii of the isomer exhibit the staggering
behavior.} \label{figrad4}
\end{figure}

\begin{figure}
\begin{center}
\resizebox{0.80\textwidth}{!}{%
  \includegraphics{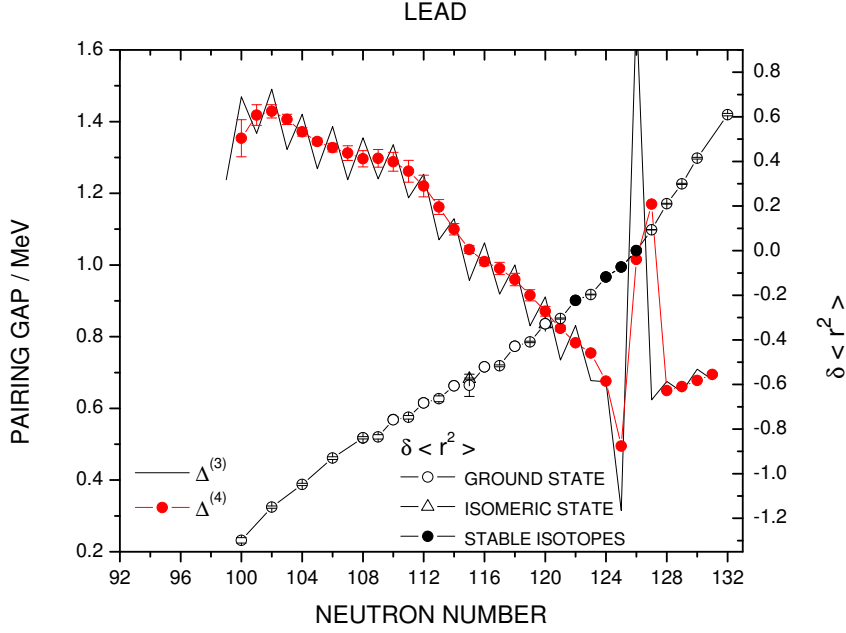}}
\end{center}
\caption{Same as Fig. \ref{figrad1} but for lead isotopes.}
\label{figrad5}
\end{figure}

Nuclear masses, as determined in this work, and nuclear charge
radii, as obtained by optical spectroscopy from isotope shift
measurements, illustrate global and collective properties of
nuclear matter. The mass is a measure of the binding energy and
the charge radii reflect static and dynamic deformations. The
nuclear system, as a function of protons and neutrons, maximizes
the binding energy according to the nuclear shape. The pairing
energy, which is about $1~\mbox{MeV}$, plays an important role in
this context and becomes visible as odd-even effect in the masses
as well as in the charge radii.\\
Due to the complex interplay of single-particle and collective
degrees of freedom, one observes in the region of the nuclear
chart investigated in this work both a spherical and deformed
shape in the same nucleus at low excitation energies. This
phenomenon, first observed at ISOLDE/CERN in the seventies by
optical spectroscopy \cite{Bonn1972,Bonn1971} is known as "shape
coexistence or shape isomerism" and nowadays an extensive bulk of
data has been collected throughout the chart of nuclei
\cite{Wood1992,Heyd1983,Juli2001,Ulm1986}. Shape coexistence is
visible in the vicinities of the shell closures
\cite{Wood1992,Heyd1983,Ulm1986}. Most clearly it is observed in
the slope of charge radii in the mercury isotopic chain: It
manifests itself in a huge odd-even shape staggering near the $N =
104$ mid-shell region \cite{Ulm1986} and a large difference in
charge radii and therefore nuclear deformation between the ground
and isomeric state in $^{185}\mbox{Hg}$ \cite{Dabk1979} (Fig.
\ref{figrad3}). Since the nuclear states of very different
configuration are (almost) degenerate in energy, one cannot expect
to observe any major trend in the slope of the masses or binding
energies as a function of neutron number. In fact, only now the
high mass resolving power $m/\Delta m \mbox{(FWHM)}$ of up to
$10^7$ of the Penning trap mass spectrometer ISOLTRAP enables one
to resolve ground and isomeric states in this region of the
nuclear chart where shape coexistence occurs. The implementation
of many new accurate mass data from ISOLTRAP
(\cite{Schw1998,Schw2001,Kohl1999} and this work) to the data base
of the Atomic Mass Evaluation and a large data set of charge radii
\cite{Klug2003,Witt2007} stimulated an attempt to reveil a
correlation between masses and charge radii \cite{Webe2005b}.\\
In order to uncover finer details in the binding energies
regarding odd-even staggering and general trends
\cite{Satu1998,Doba2001}, the absolute values of the deduced
pairing gap parameters for
neutrons are plotted instead of the experimental masses.\\
The three-point indicator for the shell gap parameter $\Delta$ is
given by
\begin{equation}
\vspace*{-0.045cm}
    \label{eq1}
    \Delta^3(N)  =  \frac{(-1)^N}{2} \left[B(N-1)+B(N+1)-2B(N)\right],
\end{equation}
which is often interpreted as a measure of the empirical pairing
gap \cite{Satu1998}. Another commonly used relation is the
four-point indicator, which averages the $\Delta^3(N)$ values and
is given by
\begin{equation}
    \label{eq2}
    \Delta^4(N) = \frac{1}{2} \left[\Delta^3(N) +
    \Delta^3(N-1)\right].
\vspace*{-0.045cm}
\end{equation}
Figures \ref{figrad1} - \ref{figrad5} compare the results from
mass measurements with those from isotope shift determinations for
the isotopic chains of platinum, gold, mercury, thallium, and
lead. For these elements, extensive data are available. In order
to allow an easy comparison of pairing gap parameters with charge
radii for the different elements, the scales of the x- and y-axis
are made identical in Figs. \ref{figrad1} - \ref{figrad5}. The
pairing gap parameters of the isotopic chains of the even-$Z$
elements lead, mercury and platinum are nearly identical in the
regions of $N < 99$ and $107 < N < 119$, respectively. The pairing
gap parameters for the isotopes of the odd-$Z$ elements thallium
and gold show a similar behavior but the pairing gap is smaller in
magnitude. Around mid-shell neutron number $N = 104$ a minimum
develops which is barely visible in the case of lead. It becomes
sharper and is shifted to $N = 105$ for thallium, distinct and
narrow for mercury
but very much pronounced and broader for gold and platinum.\\
The slopes of the isotopic chains of lead, thallium and mercury
are almost identical in the region where data are available for
the same neutron number. Only the odd-even staggering is much more
pronounced for the thallium isotopes as compared to those of lead
and mercury.\\
Down to neutron number $N = 108$, the slope of the charge radii
for the isotopes of gold and platinum is a little bit smaller than
for the other isotopic chains which indicates increasing
deformation. For $N = 105$ in the case of mercury and $N = 107$
for gold a sudden change of the charge radii is observed
which is due to a sharp change in deformation.\\
Comparing the slopes of pairing parameters with those of the
charge radii, one observes that the extremely sharp discontinuity
at the shell closure at $N = 126$ is visible in the charge radii
only as a small kink in their slope, that the pronounced increase
of deformation at mid-shell $N = 104$ is reflected as a local
minimum in the pairing parameters, and that the size of the
odd-even staggering in the pairing parameter $\Delta^3(N)$ is not
reflected in the size of the one of the charge radii. There is an
interesting observation: For thallium an unusual odd-even
staggering between the isomers below mass number $A = 193$ is
observed.\\
The new data obtained in this region of the nuclear chart near $Z
= 82$ will hopefully stimulate theoretical work to develop nuclear
models further which are able to describe simultaneously
experimental mean square charge radii as well as experimental
masses of spherical and deformed nuclei in a satisfactory way. Up
to now, only little attention has been paid to such investigations
which concern model-independent gross properties of nuclear
matter. For example, the charge radii are only poorly described by
macroscopic-microscopic mass models which are, however, quite
successful in describing nuclear masses \cite{Buch2005}. Even more
demanding is a description of the odd-even staggering of masses
and of the charge radii as attempted in Ref. \cite{Faya2001} for
spherical nuclei.
\clearpage
\clearpage
\section{\label{k57}Summary and Outlook}
The fine structure of the mass surface is important for the
elucidation of the dramatic shape changes that appear so suddenly
in this region of the nuclear chart. For most of the nuclides that
were investigated in this work a clear mass-to-state assignment
was possible. Few exceptions due to an insufficient resolving
power during the measurement are the nuclides $^{187}\mbox{Tl}$,
$^{190}\mbox{Bi}$, $^{192}\mbox{Bi}^m$, and $^{197}\mbox{Bi}$.
Spectroscopic data giving production ratios could assist these
identifications using Eq.~(\ref{54-10}). The excitation energies
of bismuth isotopes (odd-odd nuclei) suffer from particulary high
uncertainties. Hence, these cases should be addressed in a future
mass measurement requiring the following conditions: (i) If the
half-lives of the states are sufficiently long in order to allow
an extremely high-resolution measurement: A mass resolution of,
for example, $40~\mbox{keV}$ can be obtained with an excitation
time of $10~\mbox{s}$. In these cases the count-rate-class
analysis can give a meaningful interpretation, {\it i.e.} a
possible contamination is detected. (ii) Similar to the
measurements of both states of $^{187}\mbox{Pb}$ \cite{Webe2005a},
a narrow-band laser ionization with a bandwidth of
$1.2~\mbox{GHz}$ can be used to selectively ionize one of the
states. As a prerequisite, the element of interest has to be
accessible for laser ionization and the hyperfine structure
splitting of the isomeric components with different nuclear spin
has to be known. (iii) In cases where the half-life is not
sufficiently long to perform a measurement with extended
excitation time, the mass determination should be supported by
nuclear spectroscopy in order
to enable a clear identification of states.\\

\textbf{Acknowledgements}\\

We gratefully acknowledge the ISOLDE technical group and the RILIS
group for assistance during the experiments. This work was
supported by the German Ministry for Education and Research (BMBF
contracts 06MZ962I, 06MZ21S, and 06LM968), by the European
Commission (contracts HPRI-CT-1998-00018 (LSF) and
HPRI-CT-2001-50034 (NIPNET)), and by the Association of Helm-
holtz Research Centers (contract VH-NG-037).

\end{document}